\documentclass[pdflatex,sn-nature]{sn-jnl}

\usepackage{graphicx}%
\usepackage{multirow}%
\usepackage{amsmath,amssymb,amsfonts}%
\usepackage{amsthm}%
\usepackage{mathrsfs}%
\usepackage[title]{appendix}%
\usepackage{xcolor}%
\usepackage{textcomp}%
\usepackage{manyfoot}%
\usepackage{booktabs}%
\usepackage{algorithm}%
\usepackage{algorithmicx}%
\usepackage{algpseudocode}%
\usepackage{listings}%
\usepackage{placeins}
\usepackage{siunitx}
\usepackage{lineno}
\theoremstyle{thmstyleone}

\theoremstyle{thmstyletwo}%

\theoremstyle{thmstylethree}%

\begin{document}

\title[Article Title]{Diffusion MRI protocol design for reliable parameter recovery: a unified framework across biophysical models}

\author*[1]{\fnm{Ekin} \sur{Taskin}}\email{ekin.taskin@epfl.ch}

\author[1]{\fnm{Jonathan} \sur{Rafael-Patiño}}

\author[2]{\fnm{Erick Jorge} \sur{Canales-Rodríguez}}

\author[1,3,4]{\fnm{Jean-Philippe} \sur{Thiran}}

\affil*[1]{Signal Processing Laboratory 5 (LTS5), École Polytechnique Fédérale de Lausanne (EPFL), Lausanne, Switzerland}

\affil[2]{Department of Signal Theory, Networking and Communications, University of Granada, Granada, Spain}

\affil[3]{Radiology Department, Centre Hospitalier Universitaire Vaudois (CHUV) and University of Lausanne (UNIL), Switzerland}

\affil[4]{Center for Biomedical Imaging (CIBM), Lausanne, Switzerland}

\abstract{
Reliable microstructural parameter estimation in diffusion MRI (dMRI) critically depends on the acquisition protocol. Yet, common protocol design criteria may not fully reflect estimator bias, parameter constraints, and nonlinear fitting behavior in complex inverse problems. Here, we introduce Protocol Optimization via Loss Assessing Parameter Recovery (POLAR), an end-to-end framework that searches the continuous acquisition space and directly minimizes multiparameter recovery error through noisy signal simulation and the chosen model fitting. We demonstrate POLAR spanning distinct microstructural environments: the Cylinder--Zeppelin model for white matter, Neurite Exchange Imaging (NEXI) for gray matter, and Cellular Exchange Imaging (CEXI) for permeable cells. Across independent held-out parameter sets, POLAR consistently outperformed model-specific reference designs, reducing RMSE by up to $50\%$ for Cylinder–Zeppelin and $74\%$ for CEXI, while reducing NEXI exchange time bias by $70\%$. These gains across distinct biophysical models, acquisition settings, and reference design strategies demonstrate the benefit of optimizing acquisition protocols through the downstream parameter-estimation problem. POLAR provides a flexible, model-specific and estimator-aware framework for designing dMRI protocols tailored to the microstructural parameters and experimental conditions of interest.
}


\maketitle

\section{Introduction}\label{sec1}
Diffusion Magnetic Resonance Imaging (dMRI) provides a non-invasive way to probe tissue microstructure by sensitizing the MR signal to the displacement of water molecules \cite{callaghan2011translational}. Advanced biophysical signal models aim to relate the measured signal to microstructural properties such as compartment sizes, volume fractions, diffusivities, and exchange rates \cite{novikov2019quantifying,jelescu2020challenges}. These models are of particular interest because they can provide tissue-specific information relevant for characterizing microstructural alterations in health and disease \cite{alexander2019imaging, jelescu2017design}. For these models to provide meaningful microstructural information, their parameters must be recovered accurately enough to ensure that the observed differences reflect genuine tissue variation rather than the estimation error. In practice, the reliability and sensitivity of these estimates depend strongly on the protocol used to acquire the dMRI signal \cite{afzali2021sensitivity}. 

This dependence is particularly important for complex multi-parameter signal models, for which the inverse problem is often ill-conditioned. Different combinations of model parameters may produce similar diffusion signal signatures, a phenomenon known as parameter degeneracy \cite{jelescu2016degeneracy}, which can lead to ambiguous estimates and hinder the reliable quantification of model parameters. As a result, a protocol that provides substantial signal attenuation or contrast may still be suboptimal for recovering the underlying microstructural parameters. This motivates the design of acquisition protocols with respect to the microstructural parameters that will ultimately be estimated from the measured signal. Although the true parameter values are unknown before acquisition, prior biological knowledge usually provides plausible ranges over which they are expected to vary. Protocol optimization can therefore use this prior range to design measurements that better discriminate between possible parameter values and reduce the uncertainty and bias of their subsequent estimation.

A range of strategies has been proposed for dMRI acquisition protocol design. Conventional protocols are often constructed using model-specific knowledge to sample selected diffusion weightings, diffusion times, or encoding shapes. Optimal experimental design methods have been applied broadly across quantitative MRI
\cite{cercignani2006optimal,bouhrara2018fisher,zhao2018optimal,scope2022efficient} and to a variety of dMRI applications, including the selection of diffusion weightings for apparent diffusion coefficient estimation \cite{brihuega2003optimization}, gradient sampling for diffusion tensor and kurtosis imaging \cite{majumdar2011diffusion,poot2010optimal}, and acquisition design for tissue microstructure and exchange models \cite{alexander2008general,drobnjak2010optimizing,lampinen2017optimal,coelho2019optimal,lampinen2020towards,morez2023optimal,farooq2026diffusion,truffet2020evolutionary}. These methods typically construct the Fisher information matrix (FIM) to quantify parameter sensitivity and optimize a scalar design criterion, such as D-optimality, which maximizes the determinant of the FIM, or a Cramér-Rao lower bound (CRLB)-based criterion derived from its inverse \cite{pukelsheim2006optimal}. Such criteria are computationally efficient and provide a theoretically grounded connection between acquisition parameters and local signal sensitivity without requiring the specification of a particular estimator. 

A complementary information-theoretic approach is Bayesian experimental design, in which acquisition settings are optimized by accounting for uncertainty in the unknown model parameters \cite{rainforth2024modern}. In particular, expected information gain (EIG) \cite{bernardo1979expected, lindley1956measure, sebastiani2000maximum} quantifies how much a measurement is expected to reduce uncertainty about the parameters by comparing the prior and posterior distributions. This formulation has recently been applied to dMRI protocol optimization for the IMPULSED model \cite{jiang2016quantification}, where Pulsed Gradient Spin-Echo (PGSE) and Oscillating Gradient Spin-Echo (OGSE) acquisition settings were optimized under clinical scanner constraints \cite{dai2026optimizing}.

Despite their theoretical appeal, these criteria evaluate acquisitions through the likelihood rather than through the estimator that will ultimately be applied. FIM- and CRLB-based criteria characterize the local behavior of the likelihood around specified parameter values and therefore do not necessarily describe the global structure of the inverse problem. Bayesian design relaxes this locality by averaging over a prior distribution, but the result remains conditional on that prior and can be computationally demanding. Moreover, the classical CRLB specifies a theoretical lower bound on the variance of unbiased estimators, whereas estimators used in dMRI are often biased, especially under low signal-to-noise ratio (SNR), bounded parameter spaces, or ill-conditioned models. The fitting procedure itself can substantially affect achievable accuracy and precision even under a fixed, CRLB-optimized acquisition: for q-space trajectory imaging, switching between linear, weighted-linear, and iteratively reweighted linear estimators changed the median relative bias in isotropic kurtosis and the order parameter by more than an order of magnitude \cite{morez2023optimal}, consistent with earlier reports of severe weighted-least-squares bias in diffusion tensor imaging (DTI) and diffusion kurtosis imaging (DKI) parameter estimation \cite{veraart2013weighted}. Imposing constraints on the fitted tensors further improved precision across evaluated metrics at a modest cost in bias \cite{morez2023optimal}. Comparable estimator-dependent differences in accuracy, precision, and convergence reliability have also been reported for nonlinear microstructure model fitting more broadly \cite{harms2017robust}. Hence, a protocol that appears informative according to a theoretical information criterion can still lead to biased, unstable, or degenerate parameter estimates in practice.

Data-driven approaches form another family of protocol design strategies. These methods have been used to reduce densely sampled acquisitions by selecting informative measurements or shells through machine-learning-based feature selection or empirical parameter-error criteria \cite{planchuelo2024optimisation,ciceri2024framework,uhl2025explainable}. They can incorporate nonlinear parameter mappings and empirical parameter distributions. However, because the reduction is performed over a predefined candidate acquisition, the resulting protocol is necessarily restricted to the measurements represented in the initial sampling scheme. The selected protocol may also depend on the training distribution, estimator, or reference parameter maps used to guide the reduction. Some data-driven reduction methods have further incorporated parameter-estimation performance into measurement selection, choosing subsets that preserve parameter estimates obtained from a densely sampled reference acquisition \cite{ciceri2024framework,planchuelo2024optimisation}. Beyond subset selection, acquisition settings have also been optimized by minimizing errors between estimated and ground-truth model parameters. Previous work \cite{gilani2016parameter} characterized how diffusion and kurtosis tensor estimation errors in DKI depend on the b-value and gradient direction sampling scheme, and used this dependency to identify improved acquisition protocols. More recently in \cite{coelho2022reproducibility,coelho2024assessment}, multidimensional diffusion and diffusion--relaxometry protocols are optimized using the estimation error of a machine-learning-based Standard Model estimator. Together, these studies reflect an emerging shift from theoretical signal-sensitivity criteria toward evaluating acquisition protocols through the downstream parameter-estimation problem.

Building on this direction, we propose Protocol Optimization via Loss Assessing Parameter Recovery (POLAR), a unified framework for dMRI protocol design across biophysical models. Previous approaches in this direction have generally either selected measurements from a predefined candidate acquisition or optimized a particular model-estimator setting. Here, parameter recovery is formulated as a general end-to-end protocol objective: for each candidate protocol, signals are generated over biologically plausible parameter ranges, corrupted according to the specified noise model, fitted using the selected estimator, and evaluated through the discrepancy between estimated and ground-truth parameters. The optimization therefore captures practical aspects of the full inverse problem, including estimator bias, nonlinear fitting behavior, practical consequences of parameter degeneracy, and parameter constraints. The POLAR framework directly optimizes sequence-level acquisition parameters within a continuous acquisition space, while accounting for hardware and sequence constraints, and can be applied with different forward models, estimators, noise models, and parameter ranges. In this work, we evaluated POLAR for PGSE acquisitions, jointly optimizing the gradient amplitude, pulse duration, and gradient separation of each measurement,while the general formulation is not restricted to a particular encoding waveform.

We evalaute the framework across three signal models representing distinct microstructural environments: the Cylinder--Zeppelin model for white-matter axonal geometry \cite{alexander2010orientationally,dyrby2013contrast}, Neurite Exchange Imaging (NEXI) for exchange in gray matter \cite{jelescu2022neurite}, and Cellular Exchange Imaging (CEXI) for permeable spherical cells \cite{gardier2023cellular}. These experiments also span distinct acquisition regimes, allowing POLAR to be assessed under different hardware constrains. Specifically, we considered the 300 mT/m ex vivo experimental setting of the Cylinder--Zeppelin reference study \cite{dyrby2013contrast}, 500 mT/m ultra-high gradient Connectome 2.0 setting for NEXI \cite{huang2021connectome,ramos2026ultra}, and 180 mT/m setting representative of the high-performance whole-body MAGNETOM Cima.X system for CEXI. The resulting POLAR protocols were evaluated against model-specific reference designs spanning CRLB-based optimization, data-driven protocol optimization, and conventional multi-$G$ and multi-$\Delta$ acquisition strategies.

\section{Results}\label{sec2}

\begin{table}[htbp]
\caption{POLAR-optimized PGSE protocols for the three signal models. The table lists the gradient amplitude $G$, the gradient pulse duration $\delta$, the gradient separation $\Delta$, and the resulting diffusion weighting $b$. Group headings give the maximum gradient amplitude $G_{\max}$ available to the optimization and the echo time of the resulting protocol. Each protocol was optimized under the constraints and settings given in Table~\ref{tab:protocol_settings}. The corresponding reference protocols are given in Supplementary Tables~\ref{tab:supp_cylzepp}--\ref{tab:supp_cexi}.}
\label{tab:polar_protocols}
\begin{tabular*}{\textwidth}{@{\extracolsep{\fill}}
    c
    S[table-format=3.0]
    S[table-format=2.1]
    S[table-format=2.1]
    S[table-format=5.0]@{}}
\toprule
Shell & {$G$ (mT/m)} & {$\delta$ (ms)} & {$\Delta$ (ms)} & {$b$ (s/mm$^2$)} \\
\midrule
\multicolumn{5}{@{}l}{\textbf{Cylinder--Zeppelin}
  ($G_{\max} = 300$\,mT/m, TE $= 52.8$\,ms)} \\
1 & 300 & 6.5 & 9.6 & 2028 \\
2 & 154 & 10.8 & 33.1 & 5799 \\
3 & 195 & 20.6 & 23.7 & 19471 \\
\midrule
\multicolumn{5}{@{}l}{\textbf{NEXI}
  ($G_{\max} = 500$\,mT/m, TE $= 43.1$\,ms)} \\
1 & 169 & 5.0 & 8.4 & 345 \\
2 & 122 & 5.0 & 16.4 & 396 \\
3 & 236 & 5.0 & 22.1 & 2035 \\
4 & 241 & 5.0 & 21.5 & 2068 \\
5 & 497 & 5.0 & 8.0 & 2796 \\
6 & 497 & 5.0 & 12.5 & 4799 \\
7 & 497 & 5.0 & 12.9 & 4953 \\
8 & 388 & 5.0 & 21.7 & 5394 \\
\midrule
\multicolumn{5}{@{}l}{\textbf{CEXI}
  ($G_{\max} = 180$\,mT/m, TE $= 63.1$\,ms)} \\
1 & 27 & 10.2 & 34.5 & 173 \\
2 & 154 & 5.1 & 8.5 & 302 \\
3 & 36 & 19.6 & 32.5 & 910 \\
4 & 38 & 23.4 & 29.5 & 1235 \\
5 & 172 & 5.1 & 28.2 & 1460 \\
6 & 141 & 11.0 & 14.2 & 1803 \\
7 & 165 & 5.4 & 47.7 & 2618 \\
8 & 79 & 20.7 & 30.7 & 4558 \\
\bottomrule
\end{tabular*}
\end{table}

\subsection{Cylinder--Zeppelin white matter model}

\begin{figure}[!htbp]
    \centering
    \includegraphics[width=0.8\textwidth]{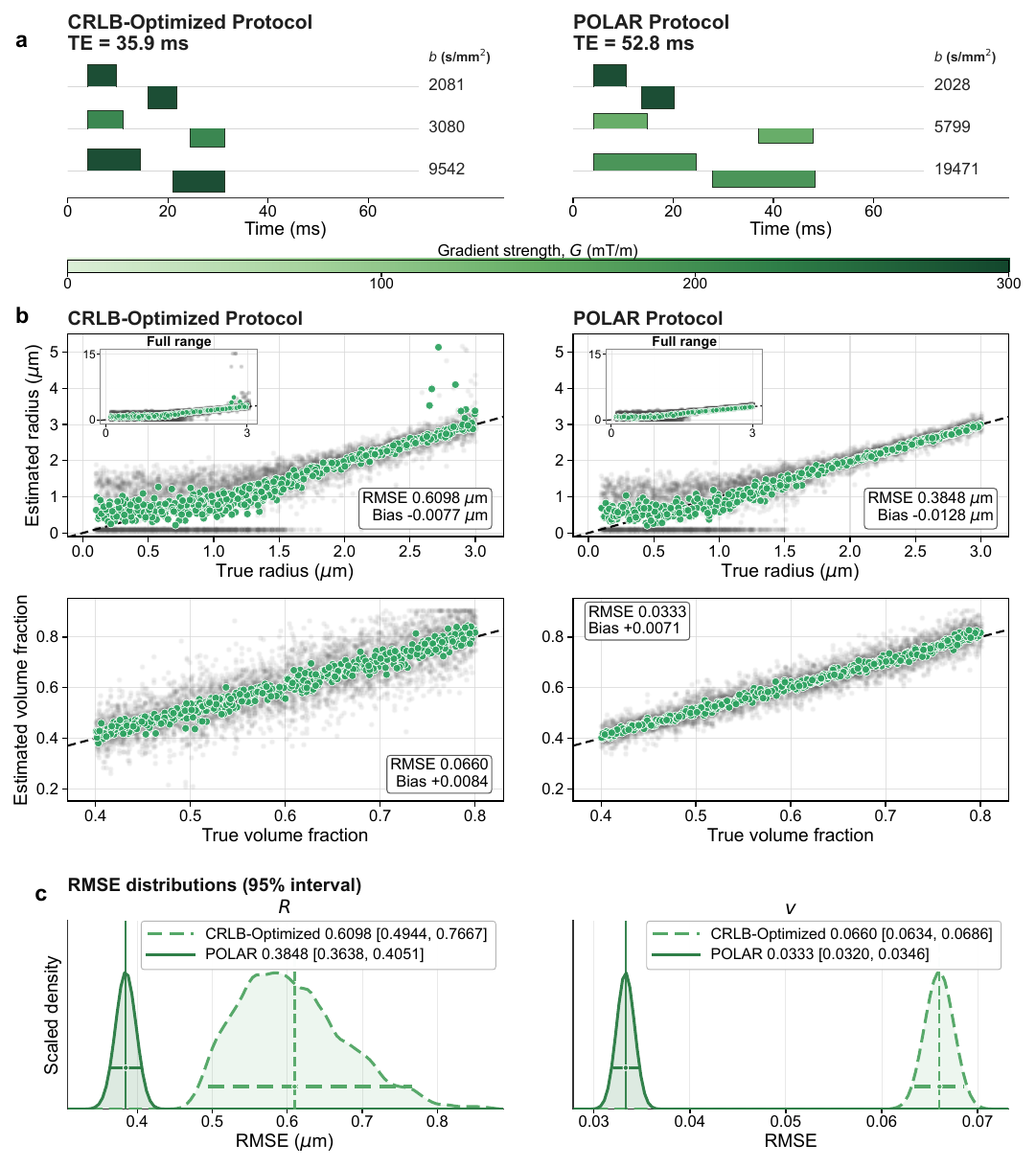}
    \caption{\textbf{Cylinder--Zeppelin CRLB-optimized and POLAR protocols, and parameter-recovery performance.} Panel (a): PGSE schematics for the CRLB-optimized and POLAR protocols. Panel (b): Estimated versus ground-truth white matter parameters for axon radius \(R\) and volume fraction \(v\) across the evaluation parameter sets. Gray points represent estimates from individual noisy realizations, while green points indicate the mean estimate for each parameter set across noise realizations. The dashed diagonal line denotes perfect agreement between the estimated and ground-truth values. Panel (c): RMSE distributions for \(R\) and \(v\) obtained from paired resampling of parameter sets and noise realizations. Dashed and solid curves correspond to the CRLB-optimized and POLAR protocols, respectively. Annotations report the RMSE together with its corresponding 95$\%$ interval.
}
    \label{fig:wm_performance}
\end{figure}

Figure \ref{fig:wm_performance} compares the CRLB-optimized and POLAR protocols and their parameter recovery performance. Relative to the CRLB-optimized protocol, the POLAR protocol selected a broader range of timing-gradient combinations and higher diffusion weightings (Table~\ref{tab:polar_protocols}, Table~\ref{tab:supp_cylzepp}). This came at the cost of a longer echo time, with $\mathrm{TE}=52.8~\mathrm{ms}$ for the POLAR protocol compared to $35.9~\mathrm{ms}$ for the CRLB-optimized protocol.

Despite the reduction in effective SNR, the POLAR protocol improved the recovery of both cylinder radius and volume fraction. Radius RMSE decreased from $0.6098~\mu\mathrm{m}$ for the CRLB-optimized protocol to $0.3848~\mu\mathrm{m}$ for the POLAR protocol, corresponding to a reduction of approximately $37\%$. The radius bias remained close to zero for both protocols ($-0.0077~\mu\mathrm{m}$ and $-0.0128~\mu\mathrm{m}$ for the CRLB-optimized and POLAR protocols, respectively). The improvement was larger for volume fraction, for which RMSE decreased from $0.0660$ to $0.0333$, corresponding to a reduction of approximately $50\%$. The POLAR protocol also produced visibly less dispersed estimates across the evaluated volume fraction range. Volume fraction bias remained small and comparable between protocols, with values of $+0.0084$ for the CRLB-optimized protocol and $+0.0071$ for the POLAR protocol.

The RMSE resampling distributions were consistent with the scatter plots and aggregate metrics (Figure \ref{fig:wm_performance}c). For both radius and volume fraction, the POLAR protocol yielded lower RMSE distributions with non-overlapping central $95\%$ resampling intervals relative to the CRLB-optimized protocol, indicating that the improvements were stable under repeated resampling of parameter sets and noise realizations.

\subsection{Neurite Exchange Imaging (NEXI)}

\begin{figure}[!htbp]
    \centering
    \includegraphics[width=\textwidth]{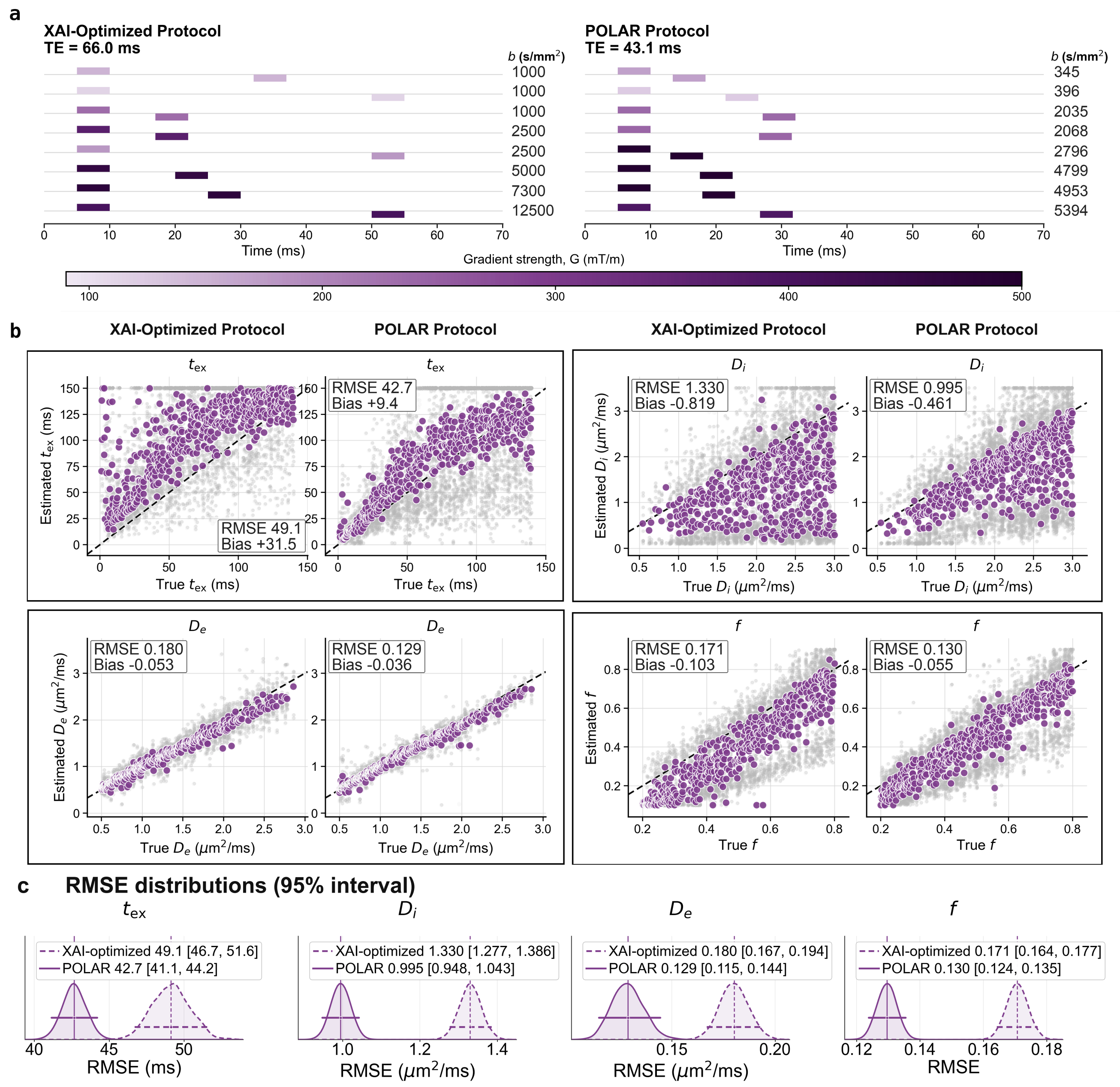}
    \caption{\textbf{NEXI parameter recovery on the continuous evaluation set.} Recovery of NEXI parameters for the XAI-optimized protocol and the POLAR protocol across 500 sampled parameter sets with 10 noise realizations per parameter set. Panel (a), timing diagrams for the XAI-optimized and POLAR protocols. Gradient pulse color encodes gradient strength $G$. Panel (b) shows estimated versus ground-truth values for exchange time $t_{ex}$, intra-neurite diffusivity $D_i$, extra-neurite diffusivity $D_e$, and intra-neurite signal fraction $f$. Gray points show individual noisy realizations, purple points show the mean estimate for each parameter set across noise realizations, and dashed black lines indicate perfect agreement between estimates and ground-truth. Panel (c) shows RMSE distributions for each parameter with 95$\%$ interval.
}
    \label{fig:nexi_performance}
\end{figure}

Figure \ref{fig:nexi_performance} compares the XAI-optimized and POLAR protocols and their corresponding parameter recovery performance. POLAR redistributed the measurements toward lower and intermediate diffusion weightings and shortened the echo time from $66.0$ to $43.1~\mathrm{ms}$ (Table~\ref{tab:polar_protocols}, Table~\ref{tab:supp_nexi}).

The POLAR protocol reduced the parameter recovery errors for all four NEXI parameters. The most notable change was observed for exchange time: RMSE decreased from $49.1$ to $42.7~\mathrm{ms}$, while the positive bias was reduced substantially from $+31.5$ to $+9.4~\mathrm{ms}$. RMSE for the intra-neurite diffusivity $D_i$ decreased from $1.330$ to $0.995~\mu\mathrm{m}^2/\mathrm{ms}$, while that of the extra-neurite diffusivity $D_e$ decreased from $0.180$ to $0.129~\mu\mathrm{m}^2/\mathrm{ms}$. Intra-neurite volume fraction RMSE decreased from $0.171$ to $0.130$. The magnitude of the bias was also reduced for $D_i$, $D_e$, and $f$.

The largest relative RMSE reduction was observed for $D_e$ ($28.3\%$), followed by $D_i$ and $f$ (approximately $25.2\%$ and $24\%$, respectively). Exchange time recovery showed a smaller relative RMSE reduction of approximately $13\%$, whereas the bias was reduced by approximately $70\%$, from $+31.5$ to $+9.4$ ms. Nevertheless, in both protocols, the estimates of exchange time $t_{ex}$ and intra-neurite diffusivity $D_i$ remained widely dispersed around the identity line, with overestimation of $t_{ex}$ and underestimation of $D_i$.

The RMSE resampling distributions were consistently shifted toward lower errors for the POLAR protocol, with non-overlapping central $95\%$ resampling intervals for all four parameters (Figure \ref{fig:nexi_performance}c).

\FloatBarrier

\subsection{Cellular Exchange Imaging (CEXI)}
\begin{figure}[htbp]
    \centering
    \includegraphics[width=\textwidth]{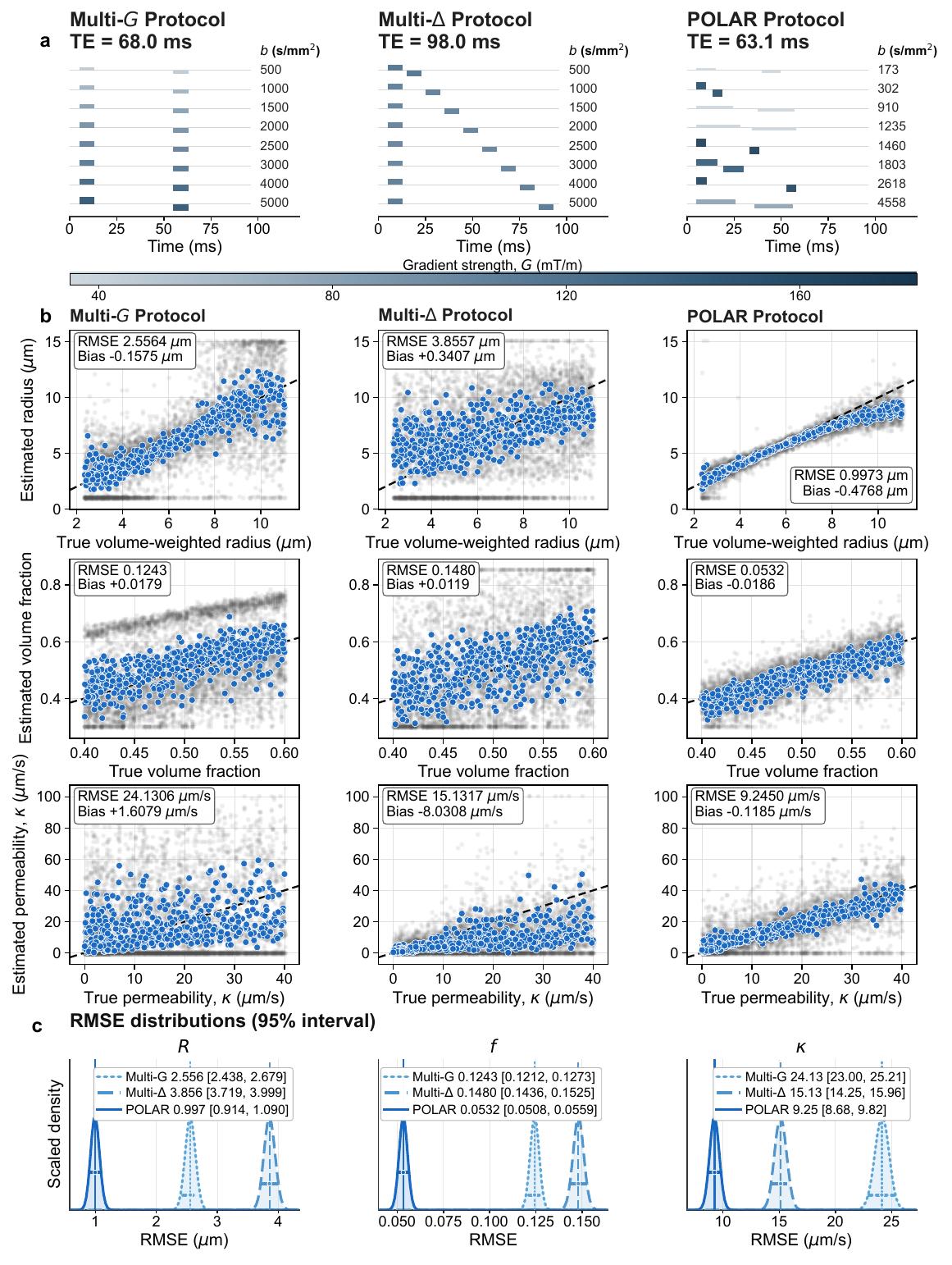}
    \caption{\textbf{CEXI multi-$G$, multi-$\Delta$ and POLAR protocols, and their parameter-recovery performance.} Panel (a): PGSE schematics for the multi-$G$, multi-$\Delta$, and POLAR protocols. Panel (b): Estimated versus ground-truth CEXI parameters for volume-weighted radius $R_{vw}$, volume fraction $f$, and permeability $\kappa$ across the evaluation parameter sets. Gray points represent estimates from individual noisy realizations, while blue points indicate the mean estimate for each parameter set across noise realizations. The dashed diagonal line denotes perfect agreement between the estimated and ground-truth values. Panel (c): RMSE distributions for $R$, $f$, and $\kappa$ obtained from paired resampling of parameter sets and noise realizations. Dotted, dashed, and solid curves correspond to the multi-$G$, multi-$\Delta$, and POLAR protocols, respectively. Annotations report the RMSE together with its corresponding 95$\%$ interval. 
}

    \label{fig:cexi_performance}
\end{figure}

Figure \ref{fig:cexi_performance} compares the reference CEXI acquisition protocols and the POLAR protocol. The multi-$G$ protocol varies the gradient strength while keeping the pulse timings fixed, whereas the multi-$\Delta$ protocol varies the gradient separation $\Delta$ with a fixed pulse duration and nearly constant gradient strength; both reference protocols used the same number of shells and covered a comparable range of diffusion weightings to POLAR (Table~\ref{tab:supp_cexi}). In contrast, the POLAR protocol jointly varies the gradient strength, pulse duration, and gradient separation, producing heterogeneous timing-gradient combinations and a nonuniform distribution of diffusion weightings (Table~\ref{tab:polar_protocols}). It also achieved the shortest echo time, at $\mathrm{TE}=63.1~\mathrm{ms}$, compared with $68.0~\mathrm{ms}$ for the multi-$G$ protocol and $98.0~\mathrm{ms}$ for the multi-$\Delta$ protocol.

The strongest improvement was observed for the volume-weighted cell radius. RMSE decreased from $2.556~\mu\mathrm{m}$ for multi-$G$ and $3.856~\mu\mathrm{m}$ for multi-$\Delta$ protocols to $0.997~\mu\mathrm{m}$ for the POLAR protocol, corresponding to reductions of approximately $61\%$ and $74\%$, respectively. 

Improvements were also observed for volume fraction and permeability. Volume fraction RMSE decreased from $0.124$ and  $0.148$ to $0.053$, while permeability RMSE decreased from $24.13$ and $15.13~\mu\mathrm{m/s}$ to $9.25~\mu\mathrm{m/s}$ for multi-$G$, multi-$\Delta$, and POLAR protocols, respectively. The POLAR protocol substantially reduced the magnitude of the permeability bias to $-0.12~\mu\mathrm{m/s}$, compared with $+1.61~\mu\mathrm{m/s}$ for the multi-$G$ protocol and $-8.03~\mu\mathrm{m/s}$ for the multi-$\Delta$ protocol.

These differences were also reflected in the paired resampling analysis, with non-overlapping central $95\%$ RMSE intervals between the POLAR and both reference protocols for radius, volume fraction, and permeability (Figure \ref{fig:cexi_performance}c).

\FloatBarrier

\section{Discussion}\label{sec12}

This study introduced POLAR, a novel unified optimization framework for diffusion MRI protocol design, and evaluated its performance using biophysical models describing distinct microstructural environments. Across the Cylinder--Zeppelin white matter model, NEXI, and CEXI, the POLAR-designed protocols consistently improved parameter recovery on independent evaluation sets. This was observed despite differences in signal models, acquisition constraints, noise conditions, estimators, and reference protocol designs. These findings suggest that optimizing protocols through the end-to-end signal simulation and fitting pipeline provides an effective strategy for acquisition design across different biophysical models, acquisition constraints, and parameter estimation settings.
Our POLAR framework evaluates candidate protocols through noisy signal simulation and model fitting, rather than relying on FIM- or CRLB-based criteria constructed from local derivatives of the analytical signal model, as commonly used in previous diffusion MRI protocol optimization studies \cite{alexander2008general,drobnjak2010optimizing,coelho2019optimal,morez2023optimal,farooq2026diffusion}. This allows protocol optimization to reflect the bias introduced by the chosen estimator and non-Gaussian noise, the effects of parameter bounds, and nonlinearities in the fitting process. Specifically, the Cylinder--Zeppelin results demonstrate the practical consequence of this distinction: the POLAR protocol produced lower empirical estimation errors than the CRLB-optimized reference under noisy, bounded nonlinear fitting.

An additional flexibility of the POLAR framework is the ability to control the relative importance of individual model parameters through the weights $w_k$ in the optimization objective (Eq. \ref{loss_function}). In the present experiments, all parameters were assigned equal weights after normalization, such that their recovery errors contributed equally to the objective. For applications in which reliable recovery of particular parameters is of primary interest, these weights can instead be adjusted to prioritize their estimation. This may be particularly useful for models containing parameters that are known to be intrinsically difficult to estimate, allowing the acquisition design to place greater emphasis on their recovery while accounting for potential trade-offs with the remaining parameters.

The POLAR framework also accounted for echo-time-dependent signal attenuation, allowing acquisition timing to influence protocol selection through its effect on SNR. For both NEXI and CEXI, the POLAR protocols achieved lower parameter estimation errors while also reducing TE relative to the data-driven and heuristic reference designs, respectively. In contrast, the Cylinder--Zeppelin results showed a different trade-off, as POLAR selected stronger diffusion weightings and a broader range of timing combinations despite the associated increase in TE and reduction in effective SNR. The resulting improvement therefore cannot be attributed to echo-time reduction alone. Together, these findings suggest that the optimization balances diffusion-encoding sensitivity against its SNR cost, with the preferred trade-off depending on the signal model and parameters of interest.

For CEXI, jointly varying gradient strength, pulse duration, and gradient separation provided a richer acquisition design than the multi-$G$ and multi-$\Delta$ reference protocols, which primarily varied one acquisition dimension. The substantial improvement in volume-weighted radius recovery suggests that heterogeneous timing-gradient combinations have improved the separation of restriction-related signal changes from those arising from volume fraction and membrane permeability. Despite the intrinsic difficulty of exchange-related parameter estimation, the POLAR protocol considerably improved permeability recovery compared to the reference protocols.

The NEXI experiment further examined whether the benefit of the POLAR protocol persisted when evaluated with a fitting procedure different from that used during protocol optimization. The final evaluation was performed using nonlinear least-squares fitting with L-BFGS-B, consistent with the fitting approach used for parameter recovery in the XAI-optimized reference study. Despite this change in parameter fitting procedure, the POLAR protocol improved recovery across all evaluated parameters and substantially reduced the systematic overestimation of exchange time. To maintain comparability with the reference design, the gradient duration $\delta$ was fixed at 5 ms, and the optimization was restricted to $G$ and $\Delta$, although the POLAR framework allows all three PGSE parameters to be optimized jointly. The POLAR protocol therefore outperformed the XAI-optimized reference protocol within a restricted acquisition space, suggesting that the improvement was not simply due to greater design flexibility or to the estimator used for inner parameter fitting during protocol optimization. Notably, the reference study found that the XAI-optimized protocol converged toward the weighted CRLB-optimized protocol design and achieved comparable performance \cite{uhl2025explainable}. The improvements observed here therefore suggest that the gains achieved by POLAR protocol can extend beyond the XAI reference to the weighted CRLB-optimized design.

The evaluation strategy also provides evidence that the observed improvements were not confined to the discrete parameter combinations used during protocol optimization. For all three models, performance was assessed on independent held-out parameter vectors sampled from continuous ranges, with the optimization grid points excluded. The lower errors on these evaluation sets therefore indicate that the POLAR-designed protocols generalized across the parameter ranges of interest rather than being tailored only to the discrete optimization grids.

For Cylinder--Zeppelin and NEXI, the number of shells was fixed to match that of the corresponding reference protocols, rather than treated as a free optimization variable. The shell-count sweep showed different saturation behavior across the three models (Figure \ref{fig:loss_vs_shell}). For Cylinder--Zeppelin, the objective loss decreased substantially up to five shells, with little additional improvement thereafter. We additionally provide the corresponding optimized five-shell POLAR protocol in Supplementary Table \ref{tab:5_shell_cyl_zepp} for future experimental evaluation. For NEXI, the loss continued to decrease across the tested range, with further gains toward nine to ten shells. CEXI similarly benefited from additional shells up to $M=9$, although the improvement from eight to nine shells was comparatively small; $M=8$ was therefore selected as a compromise between parameter recovery and acquisition complexity. Hence, the Cylinder--Zeppelin and NEXI results reported here represent controlled comparisons at matched shell count rather than the lowest losses achievable within the shell-count ranges explored.

All protocols in this work were optimized within the PGSE family, which is available on clinical and preclinical systems and is shared by the reference protocols against which POLAR was compared. The objective is nonetheless defined on the acquisition parameterization rather than on a specific sequence, so that alternative encodings require redefining only the design variables, the forward model, and the feasibility constraints. For OGSE sequences, the design variables extend the PGSE set with the oscillation frequency, or equivalently the number of gradient lobes \cite{drobnjak2016pgse}. Tensor-valued encoding introduces the b-tensor shape as a further dimension \cite{morez2023optimal}, while free waveform designs parameterize the discretized gradient trajectory subject to amplitude, slew rate, and heating limits \cite{sjolund2015constrained}. The principal cost is computational, since analytical PGSE signal expressions must be replaced by formulations derived for the chosen waveform \cite{ianucs2013gaussian} or by numerical simulation \cite{rafael2020robust}, over a substantially larger design space. Beyond extending the encoding scheme, the same end-to-end optimization formulation could also be applied to acquisition design for other downstream inference tasks, including myelin sheath radius estimation \cite{canales2025diffusion}, spherical deconvolution \cite{tournier2007robust, jeurissen2014multi, taskin2025spatially}, and axon radius estimation \cite{veraart2020noninvasive, assaf2008axcaliber, barakovic2023estimating}, by defining a task-specific recovery objective.

A limitation of the present study is that the evaluation was based on simulations, with signals generated and fitted using the analytical signal model. For CEXI, the evaluation additionally included a controlled model mismatch, because polydisperse ground-truth signals were fitted with a monodisperse effective-radius model. This setting provides exact ground truth and isolates the effect of acquisition design, but it does not assess robustness to mismatch between the assumed signal model and real tissue, or to acquisition effects such as motion or eddy-current distortions. A further limitation is that the protocols were optimized and evaluated within predefined biologically plausible parameter ranges. These ranges were selected based on prior knowledge of the tissues considered and were chosen to cover a broad range of biologically plausible values, but values outside them were not evaluated. Although the framework can readily accommodate alternative parameter ranges for healthy or diseased tissue, further experiments are needed to determine how protocols optimized within one parameter domain perform when applied beyond it. Therefore, future work should evaluate the POLAR-designed protocols in \emph{in vivo} acquisitions, and assess their robustness to model mismatch, acquisition imperfections, and parameter values beyond the optimization domain.

\section{Conclusion}\label{sec13}
We presented POLAR, a unified end-to-end framework for optimizing dMRI acquisition protocols to improve the reliability of microstructural parameter recovery. By optimizing protocols based on downstream parameter estimation performance, the optimization objective accounts for practical aspects of the inverse problem, such as estimator bias, nonlinear fitting behavior, and realistic noise model. Although demonstrated here across distinct biophysical models and acquisition settings, the POLAR framework is not tied to these specific configurations and can be applied with any chosen forward signal models, noise models, estimators, and parameter ranges, while also accommodating sequence and hardware constraints.

Across the demonstrated Cylinder--Zeppelin, NEXI, and CEXI models, the POLAR protocols improved parameter recovery relative to model-specific reference designs, including Fisher-information-based optimization, data-driven protocol reduction, and conventional acquisition schemes. These improvements were reflected in lower estimation errors and reduced parameter bias over independent evaluation sets extending beyond the discrete parameter grids used during optimization, supporting the generalization of the POLAR protocols across the parameter ranges of interest. Overall, the results demonstrate that designing acquisitions according to their parameter-recovery performance can provide advantages over criteria based only on theoretical information or selection from a predefined set of measurements.

The proposed POLAR framework provides a flexible, model-specific, and estimator-aware strategy for optimizing dMRI protocols tailored to the microstructural parameters and experimental conditions of interest.

\section{Methods}\label{sec11}

\subsection{POLAR framework}

\begin{figure}[htbp]
    \centering
    \includegraphics[width=\textwidth]{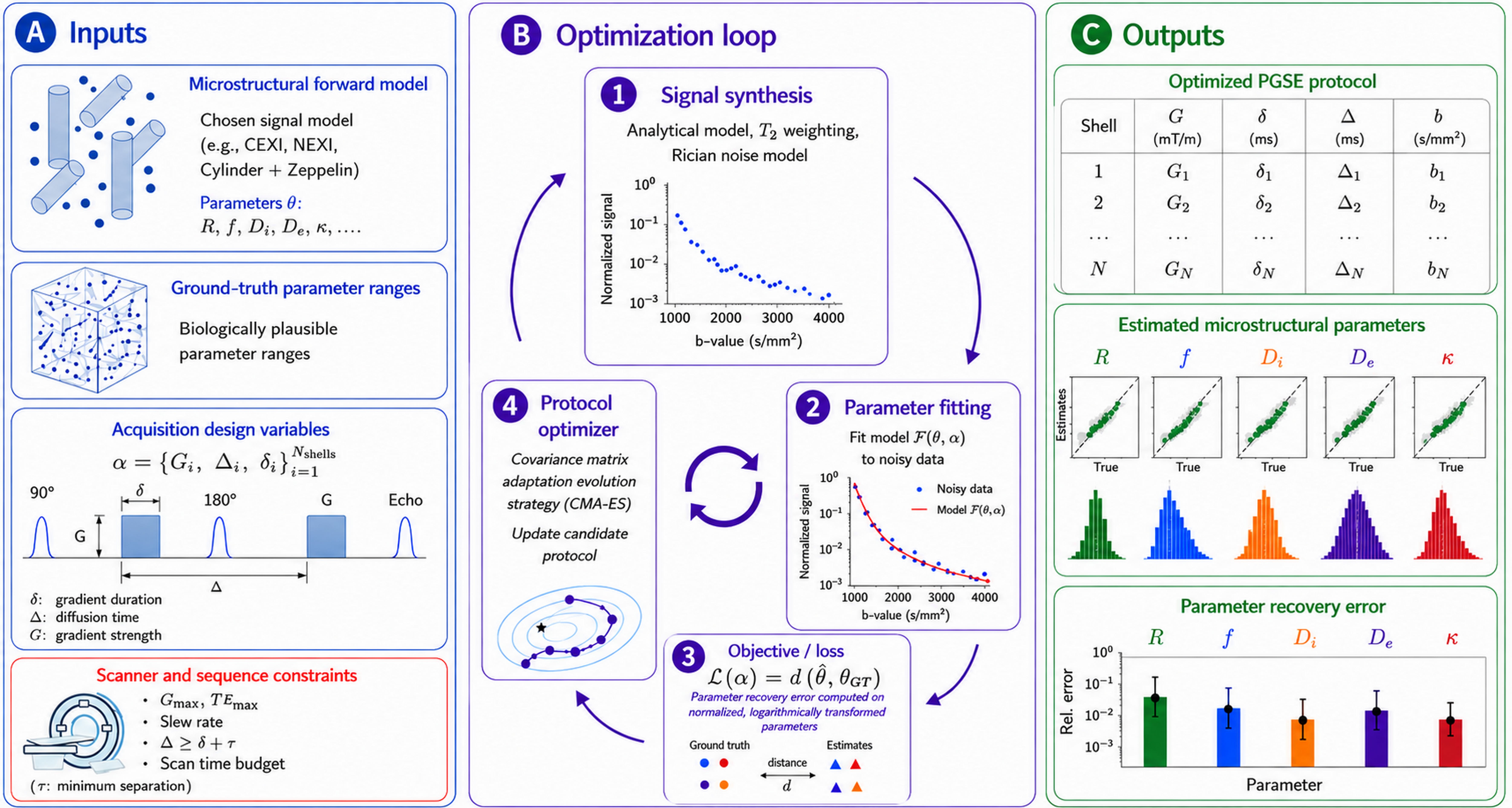}
    \caption{\textbf{POLAR framework for dMRI protocol design.} Candidate acquisition protocols are evaluated by simulating noisy diffusion-weighted signals from known ground-truth microstructural parameters, fitting the corresponding signal model, and computing a parameter-recovery loss between the estimated and true parameters. The optimizer iteratively updates the acquisition protocol under scanner and sequence constraints to improve parameter-recovery performance.
}
    \label{fig:optimization_framework}
\end{figure}

The POLAR framework optimizes diffusion MRI acquisition protocols by directly minimizing the error in recovered microstructural parameters. The procedure includes an outer protocol optimization and an inner parameter fitting step. At each outer optimization iteration, a population of candidate protocols is generated directly in the continuous acquisition space defined by the bounds and feasibility constraints in Section \ref{sec:constraints}, rather than by selecting protocols from a predefined acquisition library. For each candidate protocol, noisy diffusion-weighted signals are generated from a predefined set of ground-truth parameter vectors, and the corresponding signal model is fitted to obtain parameter estimates. The resulting discrepancy between the estimated and ground-truth parameters defines the parameter-recovery loss, which is returned to the outer optimizer to update the candidate protocol. The optimization objective therefore jointly accounts for the forward signal model, noise model, fitting procedure, and parameter constraints, while reflecting the practical consequences of parameter degeneracy. The complete POLAR workflow is summarized in Figure~\ref{fig:optimization_framework}.

Let $\boldsymbol{\alpha}=\{\boldsymbol{\alpha}_m\}_{m=1}^{M}$ denote a candidate protocol containing $M$ measurements, where $\boldsymbol{\alpha}_m$ specifies the acquisition parameters of measurement $m$. Let $\boldsymbol{\theta}_j$ denote the $j^{th}$ ground-truth parameter vector. Its entries are the microstructural parameters of the chosen signal model, such as radius, volume fractions, diffusivities, or exchange times. The corresponding noiseless diffusion signal was generated using the model-specific forward signal model $S$:

\begin{equation}
    \mathbf{s}_j(\boldsymbol{\alpha})
    =
    S\left(
        \boldsymbol{\theta}_j;
        \boldsymbol{\alpha}
    \right),
\end{equation}

To account for echo-time-dependent signal attenuation, all measurements in a candidate protocol were scaled by $T_2$ transverse relaxation time. The echo time, denoted $TE(\boldsymbol{\alpha})$, was determined by the longest echo time required among the measurements in the protocol, assuming all measurements within a candidate protocol to be acquired at a common echo time. The corresponding signal after $T_2$ relaxation was defined as
\begin{equation}
    \mathbf{s}^{T_2}_j(\boldsymbol{\alpha})
    =
    \mathbf{s}_j(\boldsymbol{\alpha})
    \exp\left[
        -\frac{\mathrm{TE}(\boldsymbol{\alpha})}{T_2}
    \right],
\end{equation}

The noise standard deviation was defined relative to the SNR at the reference echo time $\mathrm{TE}_\mathrm{ref}$:

\begin{equation}
    \sigma
    =
    \frac{\exp\left(-\mathrm{TE}_{\mathrm{ref}}/T_2\right)}
    {\mathrm{SNR}}.
\end{equation}

The noise level was fixed relative to the reference acquisition, whereas the signal amplitude depended on the echo time of each candidate protocol. Consequently, protocols requiring longer echo times were evaluated at a lower effective SNR.

Rician magnitude noise was then added independently to each simulated measurement. For noise realization $r$, the noisy signal for measurement $m$ was

\begin{equation}
    \tilde{s}_{j,r,m}(\boldsymbol{\alpha})
    =
    \sqrt{
        \left[
            s^{T_2}_{j,m}(\boldsymbol{\alpha})
            +
            \epsilon^{R}_{j,r,m}
        \right]^2
        +
        \left[
            \epsilon^{I}_{j,r,m}
        \right]^2
    },
    \qquad
    \epsilon^{R}_{j,r,m},
    \epsilon^{I}_{j,r,m}
    \overset{\mathrm{i.i.d.}}{\sim}
    \mathcal{N}(0,\sigma^2).
\end{equation}

When powder-averaged signals were used, independent noisy samples were generated in individual measurement directions and averaged before model fitting.

For each ground-truth parameter vector and noise realization, the noisy signal $\tilde{\mathbf{s}}_{j,r}(\boldsymbol{\alpha})$ was fitted using the same signal model to obtain the estimated parameter vector:

\begin{equation}
\hat{\boldsymbol{\theta}}_{j,r}(\boldsymbol{\alpha})
=
\underset{\boldsymbol{\theta}\in\Theta}{\operatorname{arg\,min}}
\;
\mathcal{C}\!\left[
\tilde{\mathbf{s}}_{j,r}(\boldsymbol{\alpha}),
S(\boldsymbol{\theta};\boldsymbol{\alpha})
\right],
\label{eq:parameter_fitting}
\end{equation}

where $\mathcal{C}$ denotes the fitting cost and $\Theta$ the model-specific parameter bounds. In the present experiments, $\mathcal{C}$ was the unweighted sum of squared signal residuals between the noisy signal and the model prediction.

Because the fitted parameters differed in physical units, numerical scale, and admissible range, each parameter was mapped to a common normalized logarithmic coordinate system, which was used for parameter fitting and for evaluating parameter-recovery errors. For parameter $k$, a parameter value $x$ was mapped according to

\begin{equation}\label{log_norm}
    z_k(x) =
    \frac{
        \log(x)-\log(\theta_k^{\min})
    }{
        \log(\theta_k^{\max})-\log(\theta_k^{\min})
    },
\end{equation}

where $\theta_k^{\min}$ and $\theta_k^{\max}$ are the fitting bounds of parameter $k$, i.e. the lowest and highest values it was allowed to take during model fitting. The fitting bounds were set per signal model from biologically plausible ranges, and are given for each model in Section~\ref{sec:signal_models}. The normalization in Eq. \ref{log_norm} maps the estimated parameters onto a common scale in $[0,1]$, so that parameters with large absolute numerical values do not dominate the objective. Its logarithmic form makes the loss reflect relative rather than absolute recovery accuracy, which matters for parameters that span orders of magnitude. 

\textbf{Implementation and parameter fitting}. During POLAR optimization, the inner parameter fitting was performed using Covariance Matrix Adaptation Evolution Strategy (CMA-ES) \cite{hansen1996adapting} as implemented in the Python package \texttt{cma} (version 4.4.0; \url{https://github.com/CMA-ES/pycma}). The fitting was carried out in the normalized logarithmic parameter coordinates defined in Eq. \ref{log_norm} and initialized at the midpoint of these bounds ($z_k=0.5$ for all fitted parameters). For all signal models, CMA-ES was run with a population size of $\lambda=16$ and an initial step size of 0.12. Each fit comprised a single CMA-ES run with no restarts. Invalid parameter candidates or non-finite model residuals were assigned a large finite objective penalty.

The POLAR loss was defined as the mean normalized squared recovery error, averaged over the $K$ fitted parameters, $N_\theta$ ground-truth parameter vectors, and $N_r$ noise realizations:

\begin{equation}\label{loss_function}
    \mathcal{L}(\boldsymbol{\alpha})
    =
    \frac{1}{N_\theta N_r \sum_k w_k}
    \sum_{j=1}^{N_\theta}
    \sum_{r=1}^{N_r}
    \sum_{k=1}^{K}
    w_k
    \left[
        z_k\left(
            \hat{\theta}_{j,r,k}(\boldsymbol{\alpha})
        \right)
        -
        z_k\left(
            \theta_{j,k}
        \right)
    \right]^2
\end{equation}

where $K$ is the number of parameters in the fitted vector $\boldsymbol{\theta}$, listed for each signal model in Table~\ref{tab:protocol_settings}. The weight $w_k$ controls the relative contribution of parameter $k$ to the loss function in the normalized log-coordinate defined by $z_k$. Equal weights, $w_k=1$, were used in all experiments. Averaging the loss over $N_r$ independent noise realizations reduces the variability of the objective due to individual noise draws and makes protocol selection less dependent on a particular noise realization. During protocol optimization, a fresh set of independent noise realizations was sampled at each CMA-ES generation.

The outer protocol optimization was also performed using CMA-ES \cite{hansen1996adapting}. Its stochastic and derivative-free formulation is well suited to nonconvex objectives for which analytical gradients are unavailable or unreliable. The POLAR objective has these characteristics, as it combines noisy signal simulation, constrained nonlinear parameter fitting, and parameter-recovery error evaluation.
The outer CMA-ES instance was initialized at the center of the feasible acquisition bounds and updated the candidate protocols under the scanner and sequence constraints described in the following subsection. The outer CMA-ES was run with a population size of $\lambda=16$ and a maximum of 1000 iterations. This population size lies above the CMA-ES default $\lambda = 4 + \lfloor 3\ln n \rfloor$ for all search spaces considered here, where $n$ is the number of optimized acquisition parameters ($n = 9$--$27$, default $\lambda = 10$--$13$) \cite{hansen2016cma}. A single population size was used across experiments so that the optimizer configuration did not vary between signal models. Larger populations also reduce the risk of convergence to local optima, at the cost of additional function evaluations \cite{hansen2016cma}. Convergence was assessed using a stagnation criterion based on the median loss across the CMA-ES population. Using the population median rather than the best loss reduces sensitivity to isolated low-loss evaluations arising from favorable noise realizations. After a minimum of 100 generations, convergence was considered reached when the reduction in median population loss between consecutive 50-generation windows fell below the estimated standard error of this reduction. The outer CMA-ES was run without a fixed random seed. The final POLAR protocol was selected as the candidate with the lowest observed parameter-recovery loss.

\subsection{PGSE protocol parameterization and constraints}\label{sec:constraints}

We used PGSE acquisitions for all optimized protocols \cite{stejskal1965spin}. For a protocol with $M$ nonzero shells, the acquisition protocol design variables were

\begin{equation}
    \boldsymbol{\alpha} =
    \left\{
        \delta_m,\Delta_m,G_m
    \right\}_{m=1}^{M},
\end{equation}

where $\delta_m$ is the gradient pulse duration, $\Delta_m$ is the gradient separation, and $G_m$ is the gradient amplitude for measurement $m$. The corresponding b-value was computed as
\begin{equation}\label{b-value}
    b_m
    =
    \left(
        \gamma G_m \delta_m
    \right)^2
    \left(
        \Delta_m-\frac{\delta_m}{3}
    \right),
\end{equation}
with the gyromagnetic ratio $\gamma = 2.675 \times 10^8~\mathrm{rad\,s^{-1}\,T^{-1}}$.

Candidate protocols were constrained to remain physically feasible. In all signal model experiments, we imposed bounds on $\delta$, $\Delta$, and $G$. In addition, a minimum separation interval of $\tau = 3$ ms was enforced between the end of the first gradient pulse and the beginning of the second gradient pulse to accommodate the $180^\circ$ refocusing pulse, satisfying $\Delta_m \geq \delta_m + \tau$ for all $m$.

All measurements in a candidate protocol were assumed to share a common echo time. Therefore, the protocol echo time was determined by the measurement that required the longest timing and was constrained as

\begin{equation}
    TE(\boldsymbol{\alpha}) =
    \max_m
    \left(
        \delta_m+\Delta_m+TE_{\mathrm{overhead}}
    \right)
    \leq TE_{\max},
\end{equation}

where $\mathrm{TE}_{\mathrm{overhead}}$ accounts for the additional sequence-dependent timing outside the diffusion gradient encoding, including readout-related delays. The dependence of $T_2$ attenuation on $\mathrm{TE}(\boldsymbol{\alpha})$ penalized protocols with longer timing requirements through a lower effective SNR in the optimization objective.

The model-specific optimization settings used in the experiments are summarized in Table~\ref{tab:protocol_settings}. For Cylinder--Zeppelin and NEXI, the number of nonzero shells $M$ was fixed to match the shell count of the corresponding reference protocols ($M=3$ and $M=8$, respectively), so that improvements could not be attributed to a larger number of measurements. CEXI had no previously optimized reference protocol to match; therefore, $M=8$ was selected from a shell-count sweep as a compromise between parameter-recovery performance and acquisition complexity, as described in Section \ref{sec:cexi}. The reference protocols for CEXI were then designed to match this shell number and b-value range.

\begin{table}[htbp]
\centering
\caption{Model-specific PGSE constraints and optimization settings used in the experiments.}
\label{tab:protocol_settings}
\begin{tabular}{lccc}
\hline
 & Cylinder--Zeppelin & NEXI & CEXI \\
\hline
Number of shells & 3 & 8 & 8 \\
$\delta$ range & 5--40 ms & fixed 5 ms & 5--40 ms \\
$\Delta$ range & 8--50 ms & 8--50 ms & 8--50 ms \\
$G$ range & 10--300 mT/m & 10--500 mT/m & 10--180 mT/m \\
$TE_{\mathrm{overhead}}$ & 8.5 ms & 16 ms & 10 ms \\
$TE_{\max}$ & 120 ms & 120 ms & 120 ms \\
Noise realizations & 10 & 10 & 10 \\
Directional measurements per shell & 30 directions & 30 averages & 20 averages \\
Parameters $\theta$ & $v,R,d_{\parallel},d_{\perp}$ & $t_{\mathrm{ex}},D_i,D_e,f$ & $R,D_i,D_e,f,\kappa$ \\
\hline
\end{tabular}
\end{table}

\subsection{Signal models and model-specific parameter sets}\label{sec:signal_models}
\begin{figure}[!htbp]
    \centering
    \includegraphics[width=\textwidth]{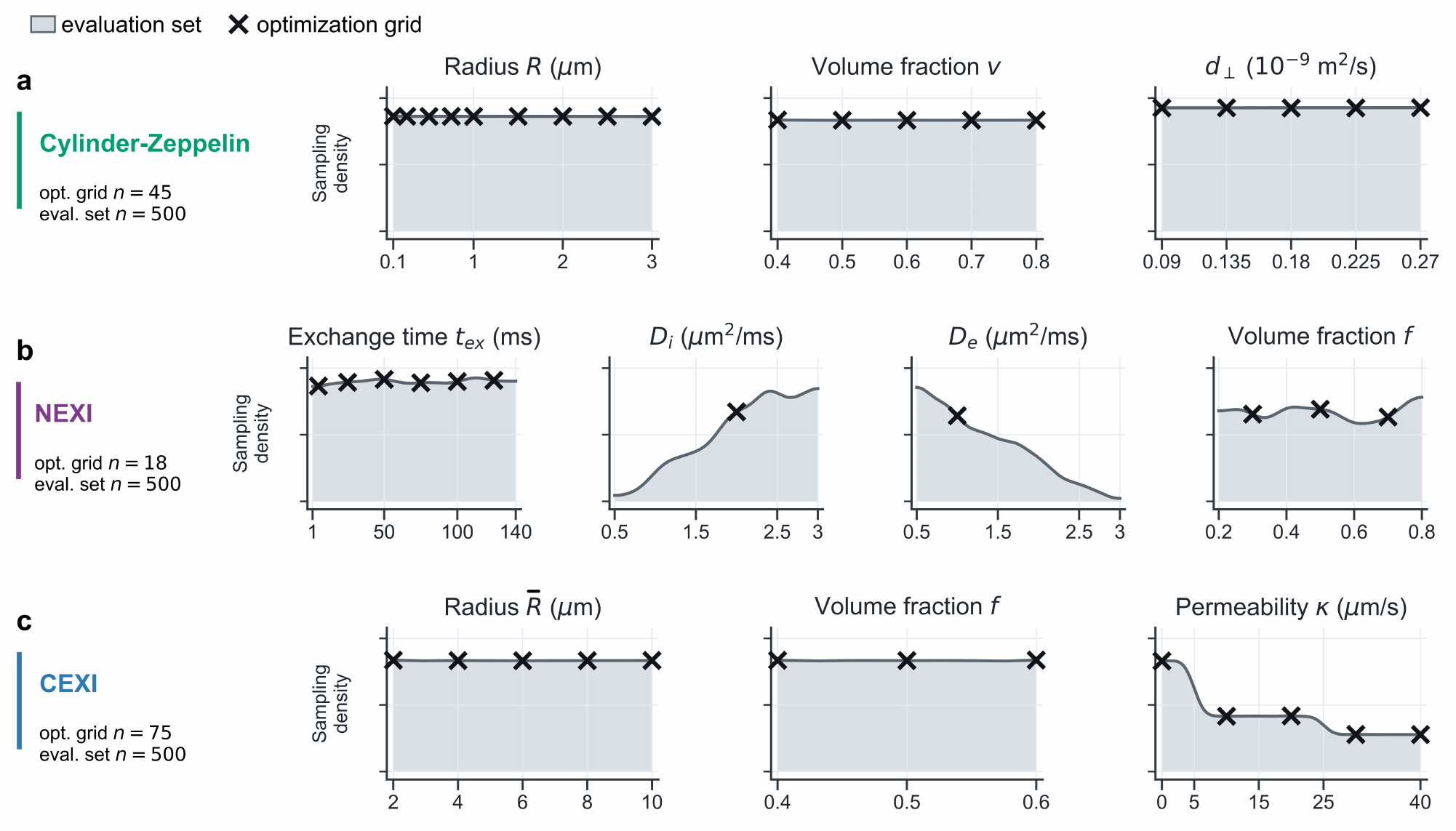}
    \caption{\textbf{Optimization and evaluation parameter sets.} Discrete optimization grids and independent held-out evaluation sets sampled from continuous parameter ranges for the Cylinder--Zeppelin, NEXI, and CEXI signal models. Black crosses indicate parameter values included in the optimization grids, while the gray filled profiles show the marginal distributions of the evaluation samples used to assess generalization beyond the optimization grids. Optimization grid points were excluded from the evaluation sets. For CEXI, the plotted radius parameter corresponds to the mean of the sphere radius distribution.
}
    \label{fig:param_grid}
\end{figure}

We evaluated the POLAR framework using three dMRI signal models representing distinct microstructural environments: the Cylinder--Zeppelin model for white matter axonal geometry, Neurite Exchange Imaging (NEXI) for exchange in gray matter, and Cellular Exchange Imaging (CEXI) for permeable spherical cells. For each signal model, two distinct parameter sets were used. First, a discrete optimization set defined the ground-truth parameter vectors over which the POLAR loss in Eq. \ref{loss_function} was evaluated during protocol optimization. The fitted parameters were not restricted to values in this discrete set, but were allowed to vary continuously within the model-specific fitting bounds. Second, an independent held-out evaluation set was sampled from continuous parameter ranges and used exclusively to evaluate parameter recovery performance beyond the discrete optimization grid (Figure \ref{fig:param_grid}). The evaluation set did not contribute to protocol selection. The evaluation sets were generated using Latin hypercube sampling \cite{ef76b040-2f28-37ba-b0c4-02ed99573416}, which stratifies each parameter range into equally represented intervals and therefore provides more systematic coverage of the marginal parameter ranges than independent random sampling at the same sample size. Parameter vectors corresponding to optimization grid points were excluded from the evaluation sets. Noise and relaxation settings were chosen separately for each model to reproduce the experimental regime of the corresponding reference study or the intended acquisition setting. 

\subsubsection{Cylinder--Zeppelin white matter model}\label{sec:cylzepp}

The white matter signal model represented tissue as a two-compartment Cylinder--Zeppelin model. The intra-axonal compartment was modeled as restricted diffusion inside parallel impermeable cylinders using the AMICO CylinderGPD model (\url{https://github.com/daducci/AMICO/}) \cite{daducci2015accelerated}, based on the Gaussian phase distribution approximation. The extra-axonal compartment was modeled as anisotropic Gaussian diffusion using the Zeppelin model. The fitted parameter vector was $
    \boldsymbol{\theta}_{\mathrm{Cyl-Zepp}}
    =
    [v,R,d_{\parallel},d_{\perp}],
$
where $v$ is the intra-axonal volume fraction, $R$ is the cylinder radius, $d_{\parallel}$ is the diffusivity parallel to the cylinder axis, and $d_{\perp}$ is the perpendicular extra-axonal diffusivity.

For measurement $m$, the total signal was

\begin{equation}
    S_m(\boldsymbol{\theta}; \boldsymbol{\alpha}_m, \phi)
        = v\, S_{cyl,m}(\phi) + (1 - v)\, S_{zepp,m}(\phi) , \label{cyl_signal}
\end{equation}

where $\phi$ is the angle between the diffusion gradient and the cylinder axis. The corresponding compartment expressions are given in Supplementary Material Section~\ref{sec:supp_cyl_zepp}. The Cylinder--Zeppelin signal was evaluated at each measured gradient direction rather than powder-averaged.

For the optimization grid, following the ex vivo white matter setting of the reference study \cite{dyrby2013contrast}, the parallel diffusivity was fixed to $d_{\parallel}=0.45\times 10^{-9}~\mathrm{m^2/s}$. This relatively low value reflects the ex vivo, room-temperature experimental conditions under which the reference protocol was derived. The perpendicular diffusivity was defined according to the tortuosity model \cite{szafer1995theoretical} as
$d_{\perp}=d_{\parallel}(1-v).$
The cylinder radius and intra-axonal volume fraction parameters were sampled from
$
    R \in \{0.1,0.25,0.5,0.75,1.0,1.5,2.0,2.5,3.0\}~\mu\mathrm{m},
    v \in \{0.4,0.5,0.6,0.7,0.8\}.
$
Although only $R$ and $v$ varied independently across the optimization grid, all four model parameters were fitted during parameter recovery. Fitting was performed using CMA-ES with the fitting bounds $v \in [0.2,0.9], R \in [0.1,15]~\mu\mathrm{m},d_{\parallel} \in [0.1,2.5]\times 10^{-9}~\mathrm{m^2/s},d_{\perp} \in [0.01,1]\times 10^{-9}~\mathrm{m^2/s}$.

The held-out evaluation set contained 500 parameter vectors sampled from $
    R \in [0.1,3.0]~\mu\mathrm{m},
    v \in [0.4,0.8],
$
with \(d_{\parallel}=0.45\times10^{-9}~\mathrm{m^2/s}\) fixed and \(d_{\perp}=d_{\parallel}(1-v)\).

The POLAR protocol contained \(M=3\) nonzero PGSE shells. Each shell was acquired along 30 diffusion directions, with an additional non-diffusion-weighted measurement. For the Cylinder--Zeppelin model optimization, the diffusion weighting of each shell was constrained to $b_{max} \leq 20,000~\mathrm{s/mm^2}$, consistent with the upper range of diffusion weightings obtained in previous experiment-design work for axon-radius estimation \cite{alexander2008general}. The POLAR protocol was compared with the CRLB-optimized protocol reported in \cite{dyrby2013contrast}, obtained using the experiment design framework of \cite{alexander2008general} for a maximum gradient strength of 300 mT/m (Table~\ref{tab:supp_cylzepp}). The same gradient constraint was therefore adopted for POLAR optimization.

Following the reference study \cite{dyrby2013contrast}, synthetic signals included $T_2$-attenuation with $T_2=50~\mathrm{ms}$, and Rician noise was generated with \(\mathrm{SNR}=20\) using \(TE=62.7~\mathrm{ms}\) as the reference signal level. Each parameter vector was evaluated using 10 independent noise realizations, with the same noise realizations used across both protocols to enable paired comparisons.

\subsubsection{Neurite Exchange Imaging}\label{sec:nexi}
 
Neurite Exchange Imaging (NEXI) models diffusion in gray matter tissue with water exchange between intra-neurite and extra-neurite compartments \cite{jelescu2022neurite}. The intra-neurite compartment was represented as randomly oriented sticks, while   extra-neurite compartment was modeled as isotropic Gaussian diffusion, with exchange between compartments described by the anisotropic K\"arger model \cite{karger1985nmr}. The fitted parameter vector was
$
    \boldsymbol{\theta}_{\mathrm{NEXI}}
    =
    [t_{\mathrm{ex}},D_i,D_e,f],
$
where \(t_{\mathrm{ex}}\) is the exchange time, \(D_i\) and \(D_e\) are the intra- and extra-neurite diffusivities, and \(f\) is the intra-neurite signal fraction. Following the NEXI parameter constraint used in the reference study \cite{uhl2025explainable}, both the optimization grid and evaluation parameter set respected \(D_i>D_e\).

The narrow-pulse powder-averaged NEXI signal model was used for all experiments. For measurement $m$,

\begin{equation}\label{nexi_signal}
    S_m(\boldsymbol{\theta}; \boldsymbol{\alpha}_m)
    = \int_0^1 \Big[ P(x)\, e^{-\eta_+(x)}
        + \big(1 - P(x)\big)\, e^{-\eta_-(x)} \Big] \, dx ,
\end{equation}

where $x$ is the cosine of the angle between the diffusion gradient and the neurite orientation. Here, $f$ is the physical intra-neurite signal fraction, whereas $P(x)$ is an angle-dependent apparent fraction in the bi-exponential signal expression. The expressions for $P(x)$ and $\eta_\pm(x)$ are given in Supplementary Material Section \ref{sec:supp_nexi}. NEXI signals were computed using the graymatter$\_$swissknife implementation of the model (\url{https://github.com/QuentinUhl/graymatter_swissknife}) \cite{uhl2025human, uhl2024quantifying}.

The optimization grid contained 18 parameter vectors defined by:
$
t_{\mathrm{ex}} \in \{5.0, 25.0, 50.0, 75.0, 100.0, 125.0\}~\mathrm{ms}$,
$D_i=2.0~\mu\mathrm{m}^2/\mathrm{ms}$,
$D_e= 1.0~\mu\mathrm{m}^2/\mathrm{ms}$,
$f \in \{0.3, 0.5, 0.7\}
$ (Figure \ref{fig:param_grid}).
All four NEXI parameters were fitted during parameter-recovery.

The held-out evaluation set contained 500 parameter vectors sampled from
$
    t_{\mathrm{ex}} \in [1,140]~\mathrm{ms},
$
$
    D_i,D_e \in [0.5,3.0]~\mu\mathrm{m}^2/\mathrm{ms},
$
$
    f \in [0.2,0.8]
$
subject to \(D_i>D_e\). Exchange time was stratified into 10 intervals to ensure coverage across its full range, and samples violating \(D_i>D_e\) were rejected. The evaluation range for $t_{ex}$ was chosen to match that used in the reference NEXI study \cite{uhl2025explainable}. The lower end of this range includes exchange times comparable to or shorter than the fixed gradient duration ($\delta=5$ ms). These samples were retained to match the reference evaluation domain and to test estimator behavior near the fast-exchange regime.

The POLAR protocol for NEXI was optimized over \(M=8\) nonzero PGSE shell-mean measurements. The gradient duration was fixed to \(\delta=5~\mathrm{ms}\), while each measurement had independently optimized diffusion time \(\Delta\) and gradient amplitude \(G\). The acquisition bounds were
$
    \Delta \in [8,50]~\mathrm{ms},$ and $
    G \in [10,500]~\mathrm{mT/m}.
$
The POLAR protocol was compared with the eight-feature NEXI protocol proposed in \cite{uhl2025explainable}, hereafter referred to as the XAI-optimized protocol. This protocol was used as a state-of-the-art NEXI-specific reference protocol from the literature. In the reference study, the XAI-optimized protocol was shown to converge to the weighted CRLB-optimized protocol. The XAI-optimized protocol was designed for the Connectome~2.0 scanner; therefore, the corresponding hardware constraints were adopted for our optimization to enable a comparable evaluation. The reference protocol was represented using the shell-mean measurements with
$
(b,\Delta)
\in
\{(1.0,12), (2.5,12), (5.0,15), (7.3,20),
(1.0,27), (1.0,45), (2.5,45), (12.5,45)\},
$
where \(b\) is in \(\mathrm{ms}/\mu\mathrm{m}^2\) and \(\Delta\) is in ms (Table~\ref{tab:supp_nexi}).

Synthetic signals included \(T_2\)-attenuation with \(T_2=60~\mathrm{ms}\). The reference echo time and noise level were chosen to match the reference NEXI protocol study on the Connectome~2.0 scanner \cite{uhl2025explainable}; therefore, Rician noise was generated with \(\mathrm{SNR}=32\) using \(TE_{\mathrm{ref}}=66~\mathrm{ms}\). For each parameter vector, the objective was averaged over 10 noise realizations, with 30 independent Rician magnitude samples averaged per shell-mean signal.

For the final performance evaluation, NEXI parameters were estimated using nonlinear least-squares (NLLS) fitting, matching the fitting approach used in the XAI reference study \cite{uhl2025explainable}. During POLAR protocol optimization, the inner parameter fitting was performed using CMA-ES, whereas the final evaluation of both the POLAR and XAI-optimized protocols used NLLS fitting. This allowed us to test whether the benefit of the POLAR protocol persisted when the inner parameter fitting problem was solved using a fitting algorithm different from that used during protocol optimization. Specifically, the sum of squared signal residuals was minimized using bound-constrained L-BFGS-B optimization with multiple random starts, implemented in SciPy \cite{virtanen2020scipy, byrd1995limited, zhu1997algorithm}. The fitting bounds were
$
    t_{\mathrm{ex}} \in [1,150]~\mathrm{ms},
    D_i,D_e \in [0.1,3.5]~\mu\mathrm{m}^2/\mathrm{ms},
    f \in [0.1,0.9].
$
Each evaluation parameter vector was simulated using 10 paired noise realizations for the XAI-optimized and POLAR protocols. 
The same noise and fitting initialization seed were used for both protocols for each parameter vector and noise index, enabling paired comparisons of recovery errors.

\subsubsection{Cellular Exchange Imaging}\label{sec:cexi}
Cellular Exchange Imaging (CEXI) models diffusion in tissue composed of permeable spherical cells \cite{gardier2023cellular}. The intracellular compartment was modeled as restricted diffusion inside spheres, the extracellular compartment as isotropic Gaussian diffusion, and exchange between compartments using a K\"arger formulation \cite{karger1985nmr}. The fitted parameter vector was $
    \boldsymbol{\theta}_{\mathrm{CEXI}}
    =
    [R,D_i,D_e,f,\kappa],
$
where \(R\) is the cell radius, \(D_i\) and \(D_e\) are the intra- and extracellular diffusivities, \(f\) is the intracellular volume fraction, and \(\kappa\) is the membrane permeability. 

CEXI signals were computed using the implementation in \url{https://github.com/jonhrafe/CEXI-public}, based on the model described in \cite{gardier2023cellular}. For measurement $m$, the diffusion-weighted signal was expressed as
\begin{equation}
    S_m(\boldsymbol{\theta}; \boldsymbol{\alpha}_m)
    = (1 - P_m)\, e^{-b_m D_{-,m}} + P_m\, e^{-b_m D_{+,m}} ,
    \label{cexi_signal}
\end{equation}
where $P_m$ is the apparent fraction and $D_{\pm,m}$ are the corresponding apparent diffusivities. Their expressions and the CEXI model quantities are provided in Supplementary Material Section~\ref{sec:supp_cexi}.

During protocol optimization and evaluation, each ground-truth parameter vector specified a Gaussian distribution of sphere radii rather than a single radius. Restricted diffusion inside a sphere of radius $R$ was described by the Gaussian phase distribution approximation $S_{\mathrm{sph}}(R; \boldsymbol{\alpha}_m)$, evaluated from the series solution of \cite{neuman1974spin}. For a set of radii $R_\ell$ drawn from the corresponding distribution, the intracellular signal and volume-weighted radius were

\begin{equation}\label{cexi_vw}
    \bar{S}_{i,m} = \frac{\sum_\ell R_\ell^{3}\,
        S_{\mathrm{sph}}(R_\ell; \boldsymbol{\alpha}_m)}{\sum_\ell R_\ell^{3}} ,
    \qquad
    R_{\mathrm{vw}} = \frac{\sum_\ell R_\ell^{4}}{\sum_\ell R_\ell^{3}} .
\end{equation}

The volume-weighted radius of the distribution was used both in the exchange rate and as the ground-truth radius against which parameter recovery was evaluated. The signal model used for fitting assumed a single radius, so a monodisperse model was fitted to polydisperse ground-truth signals.

The optimization grid used fixed diffusivities
$
    D_i = 1.0\times 10^{-9}~\mathrm{m^2/s},
    D_e = 2.0\times 10^{-9}~\mathrm{m^2/s},
$
together with five Gaussian radius distributions with mean radii of $
\bar{R} \in \{2,4,6,8,10\}~\mu\mathrm{m}$, $f \in \{0.4,0.5,0.6\}$, and
$\kappa \in \{0.01,10,20,30,40\}~\mu\mathrm{m/s}$. The corresponding standard deviations of the radius distributions were 0.50, 0.78, 0.91, 1.50, and 1.92 $\mu\mathrm{m}$, respectively. Although only \(R\), \(f\), and \(\kappa\) varied across the optimization grid, all five CEXI parameters were fitted during parameter recovery with fitting bounds $R \in [1,15]~\mu\mathrm{m},D_i \in [0.3,2.0]~\mu\mathrm{m}^2/\mathrm{ms},D_e \in [1.0,3.5]~\mu\mathrm{m}^2/\mathrm{ms},f \in [0.3,0.85], \kappa \in [0.001, 100]~\mu\mathrm{m}/\mathrm{s}$.

The held-out evaluation set contained 500 parameter vectors with mean sphere radius sampled from $ \bar{R} \in [2,10]~\mu\mathrm{m},
f \in [0.4,0.6],
\kappa \in [0.01,40]~\mu\mathrm{m/s},$
with \(D_i\) and \(D_e\) fixed as above. To ensure adequate coverage of low-permeability regimes, the permeability range was stratified into four ranges, $
\kappa \in [0.01,5], 
\kappa \in [5,15],
\kappa \in [15,25],$ and $
\kappa \in [25,40]~\mu\mathrm{m/s}$, with 125 samples drawn from each range using Latin hypercube sampling over \(\bar R\), \(f\), and \(\kappa\). Each sampled mean radius defined a Gaussian distribution of sphere radii, with its variance obtained by piecewise-linear interpolation between the radius distributions used during optimization. The corresponding volume-weighted radius $R_{vw}$ was used as the ground-truth for evaluating radius recovery.

\begin{figure}[!htbp]
    \centering
    \includegraphics[width=\textwidth]{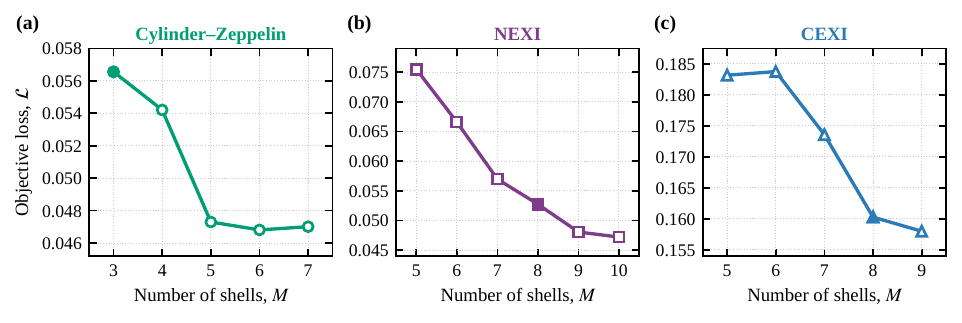}
    \caption{Effect of the number of shells $M$ on the POLAR objective loss. Objective loss $\mathcal{L}$ (Eq. \ref{loss_function}) as a function of $M$ for the (a) Cylinder--Zeppelin, (b) NEXI, and (c) CEXI signal models, obtained by independently running the POLAR optimization at each value of $M$. Filled markers indicate the number of shells used in the experiments reported in this work.}
    \label{fig:loss_vs_shell}
\end{figure}

The number of CEXI shells was selected from a sweep of $M=5, \ldots, 9$ nonzero shells, using the same CEXI objective function, parameter set, noise model, fitting procedure, and acquisition constraints (Figure~\ref{fig:loss_vs_shell}c). The objective loss decreased across the tested range up to $M=9$, with a smaller improvement from $M=8$ to $M=9$ than over the preceding shell counts. We therefore selected $M=8$ as a trade-off between parameter-recovery performance and acquisition complexity. Unlike Cylinder--Zeppelin and NEXI, no directly comparable optimized PGSE protocol was available as a reference for CEXI. We therefore constructed two conventional reference protocols that vary primarily one acquisition dimension at a time: a multi-$G$ protocol, in which diffusion weighting was varied through the gradient strength while keeping the timing parameters fixed, and a multi-$\Delta$ protocol, in which the gradient separation was varied while keeping the pulse duration fixed and the gradient strength approximately constant. Both reference protocols were designed using the same number of shells and a comparable b-value range to the POLAR protocol (Table~\ref{tab:supp_cexi}).

The POLAR protocol for CEXI contained $M=8$ nonzero PGSE shells. For protocol evaluation, each parameter vector was simulated with 10 independent noise realizations and 20 diffusion directions per shell. Synthetic signals included $T_2$ attenuation with $T_2=60~\mathrm{ms}$ and Rician noise was generated with $\mathrm{SNR}=20$, using $TE_{\mathrm{ref}}=70~\mathrm{ms}$ as the reference signal level. These values were chosen as representative settings for the CEXI acquisition considered here, rather than to reproduce a specific reference study. For each parameter vector and noise index, the same noise realization was used across protocols to enable paired comparisons.

\backmatter


\bmhead{Acknowledgements}
This work was partially supported by the Swiss National Science Foundation (SNSF) under grant number 205320$\_$204097. E.J.C.R. acknowledges financial support from the "Ramón y Cajal" Excellence Fellowship (Grant RYC2023-042763-I), funded by the Ministry of Science, Innovation and Universities (MICIU) and the Spanish State Research Agency (AEI, 10.13039/50110001103), and co-financed by the European Social Fund Plus (ESF+).

\clearpage
\section*{Supplementary Material}
 
\let\SMoldtextfraction\textfraction
\let\SMoldtopfraction\topfraction
\let\SMoldbottomfraction\bottomfraction
\let\SMoldfloatpagefraction\floatpagefraction
\renewcommand{\textfraction}{0.01}
\renewcommand{\topfraction}{0.99}
\renewcommand{\bottomfraction}{0.99}
\renewcommand{\floatpagefraction}{0.30}

\setcounter{table}{0}
\renewcommand{\thetable}{S\arabic{table}}
\setcounter{figure}{0}
\renewcommand{\thefigure}{S\arabic{figure}}

\setcounter{subsection}{0}
\renewcommand{\thesubsection}{S\arabic{subsection}}

\subsection{Reference acquisition protocols}\label{sec:supp_protocols}

This section specifies the reference acquisition protocols against which the POLAR protocols were compared. The protocols are listed in
the same convention as Table~\ref{tab:polar_protocols} of the main text.
 
\begin{table}[htbp]
\centering
\caption{Reference PGSE protocol for the Cylinder--Zeppelin model. The reference protocol is the Cramér-Rao lower bound (CRLB) optimized protocol reported in \cite{dyrby2013contrast} for a maximum gradient amplitude of 300\,mT/m, obtained using the experiment design framework of \cite{alexander2008general}. The table lists the gradient amplitude $G$, the gradient pulse duration $\delta$, the gradient separation $\Delta$, and the resulting diffusion weighting $b$. The protocol additionally includes one non-diffusion-weighted measurement. The corresponding POLAR protocol is given in Table~\ref{tab:polar_protocols} of the main text.}
\label{tab:supp_cylzepp}
\begin{tabular*}{\textwidth}{@{\extracolsep{\fill}}lcccc@{}}
\toprule
Shell & $G$ (mT/m) & $\delta$ (ms) & $\Delta$ (ms) & $b$ (s/mm$^2$) \\
\midrule
\multicolumn{5}{@{}l}{\textbf{CRLB-optimized reference} \cite{dyrby2013contrast} ($G_{\max} = 300$\,mT/m, $\mathrm{TE} = 35.9$\,ms)} \\
1 & 300 & 5.6  & 12.1 & 2081 \\
2 & 219 & 7.0  & 20.4 & 3080 \\
3 & 300 & 10.5 & 16.9 & 9542 \\
\bottomrule
\end{tabular*}
\end{table}

\begin{table}[htbp]
\centering
\caption{Reference PGSE protocol for the Neurite Exchange Imaging (NEXI) model. The reference protocol is the XAI-optimized protocol of \cite{uhl2025explainable}. Columns are as in Table~\ref{tab:supp_cylzepp}. The gradient pulse duration $\delta$ was fixed. The protocol additionally includes one non-diffusion-weighted measurement. The corresponding POLAR protocol is given in Table~\ref{tab:polar_protocols} of the main text.}
\label{tab:supp_nexi}
\begin{tabular*}{\textwidth}{@{\extracolsep{\fill}}lcccc@{}}
\toprule
Shell & $G$ (mT/m) & $\delta$ (ms) & $\Delta$ (ms) & $b$ (s/mm$^2$) \\
\midrule
\multicolumn{5}{@{}l}{\textbf{XAI-optimized reference} \cite{uhl2025explainable} ($G_{\max} = 500$\,mT/m, $\mathrm{TE} = 66.0$\,ms)} \\
1 & 233 & 5.0 & 12.0 & 1000 \\
2 & 149 & 5.0 & 27.0 & 1000 \\
3 & 114 & 5.0 & 45.0 & 1000 \\
4 & 368 & 5.0 & 12.0 & 2500 \\
5 & 180 & 5.0 & 45.0 & 2500 \\
6 & 458 & 5.0 & 15.0 & 5000 \\
7 & 472 & 5.0 & 20.0 & 7300 \\
8 & 402 & 5.0 & 45.0 & 12500 \\
\bottomrule
\end{tabular*}
\end{table}

\begin{table}[htbp]
\centering
\caption{Reference PGSE protocols for the Cellular Exchange Imaging (CEXI) model. Two conventional acquisition schemes are used as references, each varying a single encoding dimension: the Multi-$G$ scheme varies the gradient amplitude at fixed
$\delta = 8$\,ms and $\Delta = 50$\,ms, and the Multi-$\Delta$ scheme varies the gradient separation at fixed $\delta = 8$\,ms and
approximately constant $G$. Columns are as in Table~\ref{tab:supp_cylzepp}. Each protocol additionally includes one non-diffusion-weighted measurement. The corresponding POLAR protocol, which was compared against both schemes, is given in Table~\ref{tab:polar_protocols} of the main text.}
\label{tab:supp_cexi}
\begin{tabular*}{\textwidth}{@{\extracolsep{\fill}}lcccc@{}}
\toprule
Shell & $G$ (mT/m) & $\delta$ (ms) & $\Delta$ (ms) & $b$ (s/mm$^2$) \\
\midrule
\multicolumn{5}{@{}l}{\textbf{Multi-$G$ reference} ($G_{\max} = 180$\,mT/m, $\mathrm{TE} = 68.0$\,ms)} \\
1 & 48 & 8.0 & 50.0 & 500 \\
2 & 68 & 8.0 & 50.0 & 1000 \\
3 & 83 & 8.0 & 50.0 & 1500 \\
4 & 96 & 8.0 & 50.0 & 2000 \\
5 & 107 & 8.0 & 50.0 & 2500 \\
6 & 118 & 8.0 & 50.0 & 3000 \\
7 & 136 & 8.0 & 50.0 & 4000 \\
8 & 152 & 8.0 & 50.0 & 5000 \\
\midrule
\multicolumn{5}{@{}l}{\textbf{Multi-$\Delta$ reference} ($G_{\max} = 180$\,mT/m, $\mathrm{TE} = 98.0$\,ms)} \\
1 & 122 & 8.0 & 10.0 & 500 \\
2 & 112 & 8.0 & 20.0 & 1000 \\
3 & 109 & 8.0 & 30.0 & 1500 \\
4 & 108 & 8.0 & 40.0 & 2000 \\
5 & 107 & 8.0 & 50.0 & 2500 \\
6 & 107 & 8.0 & 60.0 & 3000 \\
7 & 114 & 8.0 & 70.0 & 4000 \\
8 & 119 & 8.0 & 80.0 & 5000 \\
\bottomrule
\end{tabular*}
\end{table}
 
\clearpage

 \subsection{Signal models}
 
\subsubsection{Cylinder--Zeppelin white matter model}\label{sec:supp_cyl_zepp}
The Cylinder--Zeppelin signal for measurement $m$ was defined as
\begin{equation}
    S_m(\boldsymbol{\theta}; \boldsymbol{\alpha}_m, \phi)
        = v\, S_{cyl,m}(\phi) + (1 - v)\, S_{zepp,m}(\phi) , \label{cyl_signal}
\end{equation}
where $\phi$ is the angle between the diffusion gradient and the cylinder axis.

The intra-axonal compartment was modeled as restricted diffusion inside parallel impermeable cylinders using the Gaussian phase distribution approximation. Restricted diffusion perpendicular to the cylinder axis contributes a signal attenuation that is quadratic in the perpendicular gradient amplitude, with coefficient $H_m(R, d_\parallel; \boldsymbol{\alpha}_m)$ evaluated from the series solution of van Gelderen et al. \cite{vangelderen1994evaluation}. The corresponding cylinder signal was
\begin{equation}
    S_{cyl,m}(\phi) = \exp\!\Big[ -2\gamma^2 G_m^2 \sin^2\!\phi \; H_m
        - b_m d_\parallel \cos^2\!\phi \Big] , \label{cyl_restricted}
\end{equation}

The extra-axonal compartment was modeled as anisotropic Gaussian diffusion using the Zeppelin model,

\begin{equation}
    S_{zepp,m}(\phi) = \exp\!\Big[ -b_m \big( d_\parallel \cos^2\!\phi
        + d_\perp \sin^2\!\phi \big) \Big] . \label{cyl_zeppelin} 
\end{equation}
The diffusion weighting $b_m$ was computed according to Eq. \ref{b-value} of the main text.

\subsubsection{Neurite Exchange Imaging}\label{sec:supp_nexi}
The narrow-pulse powder-averaged NEXI signal for measurement $m$ was defined as
\begin{equation}
    S_m(\boldsymbol{\theta}; \boldsymbol{\alpha}_m)
    = \int_0^1 \Big[ P(x)\, e^{-\eta_+(x)}
        + \big(1 - P(x)\big)\, e^{-\eta_-(x)} \Big] \, dx ,
\end{equation}
where $x$ is the cosine of the angle between the diffusion gradient and the neurite orientation. The physical intra-neurite signal fraction is denoted by $f$, whereas $P(x)$ is the angle-dependent apparent fraction in the bi-exponential signal expression. 

The apparent fraction and corresponding decay exponents are \cite{jelescu2022neurite}:
\begin{gather}
    P(x) = \frac{b_m \big( f D_i x^2 + (1-f) D_e \big) - \eta_-(x)}
        {\sqrt{Q(x)}} , \\
    \eta_\pm(x) = \frac{1}{2}\left[ \frac{\Delta_m}{t_{\mathrm{ex}}}
        + b_m \big( D_i x^2 + D_e \big) \pm \sqrt{Q(x)} \right] , 
\end{gather}

where $Q(x)$ is defined as
\begin{gather}
    Q(x) = \left[ (1-2f)\frac{\Delta_m}{t_{\mathrm{ex}}}
        + b_m \big( D_i x^2 - D_e \big) \right]^2
        + 4f(1-f) \left( \frac{\Delta_m}{t_{\mathrm{ex}}} \right)^2 .
\end{gather}

The quantities $P(x)$ and $\eta_\pm(x)$ therefore depend jointly on the model parameters and acquisition settings. The diffusion weighting $b_m$ was computed according to Eq.~\ref{b-value} of the main text.

\subsubsection{Cellular Exchange Imaging}\label{sec:supp_cexi}
The CEXI signal for measurement $m$ was defined as
\begin{equation}
    S_m(\boldsymbol{\theta}; \boldsymbol{\alpha}_m)
    = (1 - P_m)\, e^{-b_m D_{-,m}} + P_m\, e^{-b_m D_{+,m}} ,
    \label{cexi_signal}
\end{equation}

where the apparent fraction $P_m$ and the apparent diffusivities $D_{\pm,m}$ are defined as
\begin{gather}
    P_m = \frac{f D^{\mathrm{app}}_{i,m} + (1-f) D_e - D_{-,m}}
               {D_{+,m} - D_{-,m}} ,  \\
    D_{\pm,m} = \frac{1}{2}\left[ \tilde{D}_{i,m} + \tilde{D}_{e,m}
        \pm \sqrt{ \big( \tilde{D}_{i,m} - \tilde{D}_{e,m} \big)^{2}
        + \frac{4 k_i k_e}{q_m^{4}} } \, \right] . 
\end{gather}

The apparent intracellular diffusivity was obtained from the restricted intracellular signal $\bar{S}_{i,m}$ as $D^{\mathrm{app}}_{i,m} = -\ln \bar{S}_{i,m} / b_m$. The remaining quantities were

\begin{equation}
    q_m^2 = \frac{b_m}{\Delta_m - \delta_m/3} ,
    \qquad
    \tilde{D}_{i,m} = D^{\mathrm{app}}_{i,m} + \frac{k_i}{q_m^2} ,
    \qquad
    \tilde{D}_{e,m} = D_e + \frac{k_e}{q_m^2} .
\end{equation}

The water exchange rates from the intracellular to the extracellular compartment, $k_i$, and from the extracellular to the intracellular compartment, $k_e$, were obtained from the surface-to-volume ratio of a sphere, together with the spin conservation relation

\begin{equation}
    k_i = \frac{3\kappa}{R_{\mathrm{vw}} },
    \qquad
    k_i f = k_e (1-f) ,
\end{equation}
where $R_{\mathrm{vw}}$ is the volume-weighted radius defined in the main text.

\subsection{5-shell POLAR protocol for Cylinder--Zeppelin}\label{supp:5_shell_cyl_zepp}
The main Cylinder––Zeppelin comparison used $M=3$ nonzero shells to match the shell count of the CRLB-optimized reference protocol \cite{dyrby2013contrast}. As shown in Figure \ref{fig:loss_vs_shell}a of the main text, the POLAR objective decreased further when the number of shells was increased, with the largest additional reduction occurring up to $M=5$. Table \ref{tab:5_shell_cyl_zepp} reports the corresponding five-shell POLAR protocol obtained using the same optimization settings and acquisition constraints as the main Cylinder–-Zeppelin experiment. 

\begin{table}[htbp]
\centering
\caption{Five-shell POLAR protocol for the Cylinder-–Zeppelin model. The protocol corresponds to the $M=5$ optimization shown in Figure \ref{fig:loss_vs_shell}a of the main text. Columns give the gradient amplitude $G$, gradient pulse duration $\delta$, gradient separation $\Delta$, and resulting diffusion weighting $b$. The protocol was optimized with $G_{\max}=300$ mT/m under the same settings and constraints as the main Cylinder-–Zeppelin experiment.}
\label{tab:5_shell_cyl_zepp}
\begin{tabular*}{\textwidth}{@{\extracolsep{\fill}}lcccc@{}}
\toprule
Shell & $G$ (mT/m) & $\delta$ (ms) & $\Delta$ (ms) & $b$ (s/mm$^2$) \\
\midrule
\multicolumn{5}{@{}l}{\textbf{POLAR-optimized protocol} ($G_{\max} = 300$\,mT/m, $\mathrm{TE} = 60.2$\,ms)} \\
1 & 66  & 22.9 & 26.3 & 3071 \\
2 & 256 & 5.2 & 38.2 & 4542 \\
3 & 139 & 17.6 & 33.9 & 12063 \\
4 & 299 & 14.6 & 17.9 & 17732 \\
5 & 157 & 22.1 & 29.6 & 19192 \\
\bottomrule
\end{tabular*}
\end{table}

%

\let\textfraction\SMoldtextfraction
\let\topfraction\SMoldtopfraction
\let\bottomfraction\SMoldbottomfraction
\let\floatpagefraction\SMoldfloatpagefraction

\noindent

\newpage
\bibliography{sn-bibliography}

@article{alexander2008general,
  title={A general framework for experiment design in diffusion MRI and its application in measuring direct tissue-microstructure features},
  author={Alexander, Daniel C},
  journal={Magnetic Resonance in Medicine: An Official Journal of the International Society for Magnetic Resonance in Medicine},
  volume={60},
  number={2},
  pages={439--448},
  year={2008},
  publisher={Wiley Online Library}
}

@article{cercignani2006optimal,
  title={Optimal acquisition schemes for in vivo quantitative magnetization transfer MRI},
  author={Cercignani, Mara and Alexander, Daniel C},
  journal={Magnetic Resonance in Medicine: An Official Journal of the International Society for Magnetic Resonance in Medicine},
  volume={56},
  number={4},
  pages={803--810},
  year={2006},
  publisher={Wiley Online Library}
}

@article{lampinen2020towards,
  title={Towards unconstrained compartment modeling in white matter using diffusion-relaxation MRI with tensor-valued diffusion encoding},
  author={Lampinen, Bj{\"o}rn and Szczepankiewicz, Filip and M{\aa}rtensson, Johan and van Westen, Danielle and Hansson, Oskar and Westin, Carl-Fredrik and Nilsson, Markus},
  journal={Magnetic resonance in medicine},
  volume={84},
  number={3},
  pages={1605--1623},
  year={2020},
  publisher={Wiley Online Library}
}

@article{morez2023optimal,
  title={Optimal experimental design and estimation for q-space trajectory imaging},
  author={Morez, Jan and Szczepankiewicz, Filip and den Dekker, Arnold J and Vanhevel, Floris and Sijbers, Jan and Jeurissen, Ben},
  journal={Human Brain Mapping},
  volume={44},
  number={4},
  pages={1793--1809},
  year={2023},
  publisher={Wiley Online Library}
}

@article{farooq2026diffusion,
  title={Diffusion MRI experimental design optimization for microstructure imaging},
  author={Farooq, Hamza and Chen, Yongxin and Rasool, Ghulam and Georgiou, Tryphon and Lenglet, Christophe},
  journal={Communications Biology},
  year={2026},
  publisher={Nature Publishing Group UK London}
}

@inproceedings{coelho2019optimal,
  title={Optimal experimental design for biophysical modelling in multidimensional diffusion MRI},
  author={Coelho, Santiago and Pozo, Jose M and Jespersen, Sune N and Frangi, Alejandro F},
  booktitle={International Conference on Medical Image Computing and Computer-Assisted Intervention},
  pages={617--625},
  year={2019},
  organization={Springer}
}

@article{bouhrara2018fisher,
  title={Fisher information and Cram{\'e}r-Rao lower bound for experimental design in parallel imaging},
  author={Bouhrara, Mustapha and Spencer, Richard G},
  journal={Magnetic resonance in medicine},
  volume={79},
  number={6},
  pages={3249--3255},
  year={2018},
  publisher={Wiley Online Library}
}

@article{jelescu2016degeneracy,
  title={Degeneracy in model parameter estimation for multi-compartmental diffusion in neuronal tissue},
  author={Jelescu, Ileana O and Veraart, Jelle and Fieremans, Els and Novikov, Dmitry S},
  journal={NMR in Biomedicine},
  volume={29},
  number={1},
  pages={33--47},
  year={2016},
  publisher={Wiley Online Library}
}

@article{gardier2023cellular,
  title={Cellular Exchange Imaging (CEXI): Evaluation of a diffusion model including water exchange in cells using numerical phantoms of permeable spheres},
  author={Gardier, R{\'e}my and Villarreal Haro, Juan Luis and Canales-Rodr{\'\i}guez, Erick J and Jelescu, Ileana O and Girard, Gabriel and Rafael-Pati{\~n}o, Jonathan and Thiran, Jean-Philippe},
  journal={Magnetic Resonance in Medicine},
  volume={90},
  number={4},
  pages={1625--1640},
  year={2023},
  publisher={Wiley Online Library}
}

@article{jelescu2022neurite,
  title={Neurite Exchange Imaging (NEXI): A minimal model of diffusion in gray matter with inter-compartment water exchange},
  author={Jelescu, Ileana O and de Skowronski, Alexandre and Geffroy, Fran{\c{c}}oise and Palombo, Marco and Novikov, Dmitry S},
  journal={NeuroImage},
  volume={256},
  pages={119277},
  year={2022},
  publisher={Elsevier}
}

@article{alexander2010orientationally,
  title={Orientationally invariant indices of axon diameter and density from diffusion MRI},
  author={Alexander, Daniel C and Hubbard, Penny L and Hall, Matt G and Moore, Elizabeth A and Ptito, Maurice and Parker, Geoff JM and Dyrby, Tim B},
  journal={Neuroimage},
  volume={52},
  number={4},
  pages={1374--1389},
  year={2010},
  publisher={Elsevier}
}

@inproceedings{hansen1996adapting,
  title={Adapting arbitrary normal mutation distributions in evolution strategies: The covariance matrix adaptation},
  author={Hansen, Nikolaus and Ostermeier, Andreas},
  booktitle={Proceedings of IEEE international conference on evolutionary computation},
  pages={312--317},
  year={1996},
  organization={IEEE}
}

@article{ef76b040-2f28-37ba-b0c4-02ed99573416,
 ISSN = {00401706},
 URL = {http://www.jstor.org/stable/1268522},
 author = {M. D. McKay and R. J. Beckman and W. J. Conover},
 journal = {Technometrics},
 number = {2},
 pages = {239--245},
 publisher = {[Taylor & Francis, Ltd., American Statistical Association, American Society for Quality]},
 title = {A Comparison of Three Methods for Selecting Values of Input Variables in the Analysis of Output from a Computer Code},
 urldate = {2026-06-28},
 volume = {21},
 year = {1979}
}

@article{daducci2015accelerated,
  title={Accelerated microstructure imaging via convex optimization (AMICO) from diffusion MRI data},
  author={Daducci, Alessandro and Canales-Rodr{\'\i}guez, Erick J and Zhang, Hui and Dyrby, Tim B and Alexander, Daniel C and Thiran, Jean-Philippe},
  journal={Neuroimage},
  volume={105},
  pages={32--44},
  year={2015},
  publisher={Elsevier}
}

@article{szafer1995theoretical,
  title={Theoretical model for water diffusion in tissues},
  author={Szafer, Aaron and Zhong, Jianhui and Gore, John C},
  journal={Magnetic resonance in medicine},
  volume={33},
  number={5},
  pages={697--712},
  year={1995},
  publisher={Wiley Online Library}
}

@article{virtanen2020scipy,
  title={SciPy 1.0: fundamental algorithms for scientific computing in Python},
  author={Virtanen, Pauli and Gommers, Ralf and Oliphant, Travis E and Haberland, Matt and Reddy, Tyler and Cournapeau, David and Burovski, Evgeni and Peterson, Pearu and Weckesser, Warren and Bright, Jonathan and others},
  journal={Nature methods},
  volume={17},
  number={3},
  pages={261--272},
  year={2020},
  publisher={Nature Publishing Group US New York}
}

@article{byrd1995limited,
  title={A limited memory algorithm for bound constrained optimization},
  author={Byrd, Richard H and Lu, Peihuang and Nocedal, Jorge and Zhu, Ciyou},
  journal={SIAM Journal on scientific computing},
  volume={16},
  number={5},
  pages={1190--1208},
  year={1995},
  publisher={SIAM}
}

@article{zhu1997algorithm,
  title={Algorithm 778: L-BFGS-B: Fortran subroutines for large-scale bound-constrained optimization},
  author={Zhu, Ciyou and Byrd, Richard H and Lu, Peihuang and Nocedal, Jorge},
  journal={ACM Transactions on mathematical software (TOMS)},
  volume={23},
  number={4},
  pages={550--560},
  year={1997},
  publisher={ACM New York, NY, USA}
}

@article{dyrby2013contrast,
  title={Contrast and stability of the axon diameter index from microstructure imaging with diffusion MRI},
  author={Dyrby, Tim B and S{\o}gaard, Lise V and Hall, Matt G and Ptito, Maurice and Alexander, Daniel C},
  journal={Magnetic resonance in medicine},
  volume={70},
  number={3},
  pages={711--721},
  year={2013},
  publisher={Wiley Online Library}
}

@article{planchuelo2024optimisation,
  title={Optimisation of quantitative brain diffusion-relaxation MRI acquisition protocols with physics-informed machine learning},
  author={Planchuelo-G{\'o}mez, {\'A}lvaro and Descoteaux, Maxime and Larochelle, Hugo and Hutter, Jana and Jones, Derek K and Tax, Chantal MW},
  journal={Medical Image Analysis},
  volume={94},
  pages={103134},
  year={2024},
  publisher={Elsevier}
}

@article{ciceri2024framework,
  title={A framework for optimizing the acquisition protocol multishell diffusion-weighted imaging for multimodel assessment},
  author={Ciceri, Tommaso and De Luca, Alberto and Agarwal, Nivedita and Arrigoni, Filippo and Peruzzo, Denis},
  journal={NMR in Biomedicine},
  volume={37},
  number={8},
  pages={e5141},
  year={2024},
  publisher={Wiley Online Library}
}

@article{coelho2022reproducibility,
  title={Reproducibility of the Standard Model of diffusion in white matter on clinical MRI systems},
  author={Coelho, Santiago and Baete, Steven H and Lemberskiy, Gregory and Ades-Aron, Benjamin and Barrol, Genevieve and Veraart, Jelle and Novikov, Dmitry S and Fieremans, Els},
  journal={NeuroImage},
  volume={257},
  pages={119290},
  year={2022},
  publisher={Elsevier}
}

@article{drobnjak2010optimizing,
  title={Optimizing gradient waveforms for microstructure sensitivity in diffusion-weighted MR},
  author={Drobnjak, Ivana and Siow, Bernard and Alexander, Daniel C},
  journal={Journal of Magnetic Resonance},
  volume={206},
  number={1},
  pages={41--51},
  year={2010},
  publisher={Elsevier}
}

@article{alexander2019imaging,
  title={Imaging brain microstructure with diffusion MRI: practicality and applications},
  author={Alexander, Daniel C and Dyrby, Tim B and Nilsson, Markus and Zhang, Hui},
  journal={NMR in Biomedicine},
  volume={32},
  number={4},
  pages={e3841},
  year={2019},
  publisher={Wiley Online Library}
}

@article{afzali2021sensitivity,
  title={The sensitivity of diffusion MRI to microstructural properties and experimental factors},
  author={Afzali, Maryam and Pieciak, Tomasz and Newman, Sharlene and Garyfallidis, Eleftherios and {\"O}zarslan, Evren and Cheng, Hu and Jones, Derek K},
  journal={Journal of Neuroscience Methods},
  volume={347},
  pages={108951},
  year={2021},
  publisher={Elsevier}
}

@article{jelescu2017design,
  title={Design and validation of diffusion MRI models of white matter},
  author={Jelescu, Ileana O and Budde, Matthew D},
  journal={Frontiers in physics},
  volume={5},
  pages={61},
  year={2017},
  publisher={Frontiers Media SA}
}

@article{coelho2024assessment,
  title={Assessment of precision and accuracy of brain white matter microstructure using combined diffusion MRI and relaxometry},
  author={Coelho, Santiago and Liao, Ying and Szczepankiewicz, Filip and Veraart, Jelle and Chung, Sohae and Lui, Yvonne W and Novikov, Dmitry S and Fieremans, Els},
  journal={Human Brain Mapping},
  volume={45},
  number={9},
  pages={e26725},
  year={2024},
  publisher={Wiley Online Library}
}

@article{brihuega2003optimization,
  title={Optimization of diffusion measurements using Cramer-Rao lower bound theory and its application to articular cartilage},
  author={Brihuega-Moreno, Oscar and Heese, Frank P and Hall, Laurance D},
  journal={Magnetic Resonance in Medicine: An Official Journal of the International Society for Magnetic Resonance in Medicine},
  volume={50},
  number={5},
  pages={1069--1076},
  year={2003},
  publisher={Wiley Online Library}
}

@article{poot2010optimal,
  title={Optimal experimental design for diffusion kurtosis imaging},
  author={Poot, Dirk HJ and Den Dekker, Arnold J and Achten, Eric and Verhoye, Marleen and Sijbers, Jan},
  journal={IEEE transactions on medical imaging},
  volume={29},
  number={3},
  pages={819--829},
  year={2010},
  publisher={IEEE}
}

@article{majumdar2011diffusion,
  title={A diffusion gradient optimization framework for spinal cord diffusion tensor imaging},
  author={Majumdar, Shantanu and Zhu, David C and Udpa, Satish S and Raguin, L Guy},
  journal={Magnetic resonance imaging},
  volume={29},
  number={6},
  pages={789--804},
  year={2011},
  publisher={Elsevier}
}

@article{lampinen2017optimal,
  title={Optimal experimental design for filter exchange imaging: Apparent exchange rate measurements in the healthy brain and in intracranial tumors},
  author={Lampinen, Bj{\"o}rn and Szczepankiewicz, Filip and van Westen, Danielle and Englund, Elisabet and C Sundgren, Pia and L{\"a}tt, Jimmy and St{\aa}hlberg, Freddy and Nilsson, Markus},
  journal={Magnetic resonance in medicine},
  volume={77},
  number={3},
  pages={1104--1114},
  year={2017},
  publisher={Wiley Online Library}
}

@article{novikov2019quantifying,
  title={Quantifying brain microstructure with diffusion MRI: Theory and parameter estimation},
  author={Novikov, Dmitry S and Fieremans, Els and Jespersen, Sune N and Kiselev, Valerij G},
  journal={NMR in Biomedicine},
  volume={32},
  number={4},
  pages={e3998},
  year={2019},
  publisher={Wiley Online Library}
}

@article{jelescu2020challenges,
  title={Challenges for biophysical modeling of microstructure},
  author={Jelescu, Ileana O and Palombo, Marco and Bagnato, Francesca and Schilling, Kurt G},
  journal={Journal of neuroscience methods},
  volume={344},
  pages={108861},
  year={2020},
  publisher={Elsevier}
}

@book{pukelsheim2006optimal,
  title={Optimal design of experiments},
  author={Pukelsheim, Friedrich},
  year={2006},
  publisher={SIAM}
}

@book{callaghan2011translational,
  title={Translational dynamics and magnetic resonance: principles of pulsed gradient spin echo NMR},
  author={Callaghan, Paul T},
  year={2011},
  publisher={Oxford University Press}
}

@article{stejskal1965spin,
  title={Spin diffusion measurements: spin echoes in the presence of a time-dependent field gradient},
  author={Stejskal, Edward O and Tanner, John E},
  journal={The journal of chemical physics},
  volume={42},
  number={1},
  pages={288--292},
  year={1965},
  publisher={American Institute of Physics}
}

@article{harms2017robust,
  title={Robust and fast nonlinear optimization of diffusion MRI microstructure models},
  author={Harms, Robbert L and Fritz, Francisco J and Tobisch, Alexandra and Goebel, Rainer and Roebroeck, Alard},
  journal={Neuroimage},
  volume={155},
  pages={82--96},
  year={2017},
  publisher={Elsevier}
}

@article{gilani2016parameter,
  title={Parameter estimation error dependency on the acquisition protocol in diffusion kurtosis imaging},
  author={Gilani, Nima and Malcolm, Paul N and Johnson, Glyn},
  journal={Applied magnetic resonance},
  volume={47},
  number={11},
  pages={1229--1238},
  year={2016},
  publisher={Springer}
}

@article{veraart2013weighted,
  title={Weighted linear least squares estimation of diffusion MRI parameters: strengths, limitations, and pitfalls},
  author={Veraart, Jelle and Sijbers, Jan and Sunaert, Stefan and Leemans, Alexander and Jeurissen, Ben},
  journal={Neuroimage},
  volume={81},
  pages={335--346},
  year={2013},
  publisher={Elsevier}
}

@article{drobnjak2016pgse,
  title={PGSE, OGSE, and sensitivity to axon diameter in diffusion MRI: Insight from a simulation study},
  author={Drobnjak, Ivana and Zhang, Hui and Ianu{\c{s}}, Andrada and Kaden, Enrico and Alexander, Daniel C},
  journal={Magnetic resonance in medicine},
  volume={75},
  number={2},
  pages={688--700},
  year={2016},
  publisher={Wiley Online Library}
}

@article{ianucs2013gaussian,
  title={Gaussian phase distribution approximations for oscillating gradient spin echo diffusion MRI},
  author={Ianu{\c{s}}, Andrada and Siow, Bernard and Drobnjak, Ivana and Zhang, Hui and Alexander, Daniel C},
  journal={Journal of magnetic resonance},
  volume={227},
  pages={25--34},
  year={2013},
  publisher={Elsevier}
}

@article{sjolund2015constrained,
  title={Constrained optimization of gradient waveforms for generalized diffusion encoding},
  author={Sj{\"o}lund, Jens and Szczepankiewicz, Filip and Nilsson, Markus and Topgaard, Daniel and Westin, Carl-Fredrik and Knutsson, Hans},
  journal={Journal of magnetic resonance},
  volume={261},
  pages={157--168},
  year={2015},
  publisher={Elsevier}
}

@inproceedings{truffet2020evolutionary,
  title={An evolutionary framework for microstructure-sensitive generalized diffusion gradient waveforms},
  author={Truffet, Rapha{\"e}l and Rafael-Patino, Jonathan and Girard, Gabriel and Pizzolato, Marco and Barillot, Christian and Thiran, Jean-Philippe and Caruyer, Emmanuel},
  booktitle={International Conference on Medical Image Computing and Computer-Assisted Intervention},
  pages={94--103},
  year={2020},
  organization={Springer}
}

@article{hansen2016cma,
  title={The CMA evolution strategy: A tutorial},
  author={Hansen, Nikolaus},
  journal={arXiv preprint arXiv:1604.00772},
  year={2016}
}

@article{uhl2025human,
  title={Human gray matter microstructure mapped using neurite exchange imaging (NEXI) on a clinical scanner},
  author={Uhl, Quentin and Pavan, Tommaso and Feiweier, Thorsten and Piredda, Gian Franco and Jelescu, Ileana},
  journal={Imaging Neuroscience},
  volume={3},
  pages={IMAG--a},
  year={2025},
  publisher={MIT Press 255 Main Street, 9th Floor, Cambridge, Massachusetts 02142, USA~…}
}

@article{uhl2024quantifying,
  title={Quantifying human gray matter microstructure using neurite exchange imaging (NEXI) and 300 mT/m gradients},
  author={Uhl, Quentin and Pavan, Tommaso and Molendowska, Malwina and Jones, Derek K and Palombo, Marco and Jelescu, Ileana Ozana},
  journal={Imaging Neuroscience},
  volume={2},
  pages={imag--2},
  year={2024},
  publisher={MIT Press One Broadway, 12th Floor, Cambridge, Massachusetts 02142, USA~…}
}

@article{neuman1974spin,
  title={Spin echo of spins diffusing in a bounded medium},
  author={Neuman, CH},
  journal={The Journal of Chemical Physics},
  volume={60},
  number={11},
  pages={4508--4511},
  year={1974},
  publisher={American Institute of Physics}
}

@article{vangelderen1994evaluation,
  title={Evaluation of restricted diffusion in cylinders. Phosphocreatine in rabbit leg muscle},
  author={Vangelderen, P and DesPres, D and Vanzijl, PCM and Moonen, CTW},
  journal={Journal of Magnetic Resonance, Series B},
  volume={103},
  number={3},
  pages={255--260},
  year={1994},
  publisher={Elsevier}
}

@article{rainforth2024modern,
  title={Modern Bayesian experimental design},
  author={Rainforth, Tom and Foster, Adam and Ivanova, Desi R and Bickford Smith, Freddie},
  journal={Statistical Science},
  volume={39},
  number={1},
  pages={100--114},
  year={2024},
  publisher={Institute of Mathematical Statistics}
}

@article{bernardo1979expected,
  title={Expected information as expected utility},
  author={Bernardo, Jos{\'e} M},
  journal={the Annals of Statistics},
  pages={686--690},
  year={1979},
  publisher={JSTOR}
}

@article{lindley1956measure,
  title={On a measure of the information provided by an experiment},
  author={Lindley, Dennis V},
  journal={The Annals of Mathematical Statistics},
  volume={27},
  number={4},
  pages={986--1005},
  year={1956},
  publisher={Institute of Mathematical Statistics}
}

@article{sebastiani2000maximum,
  title={Maximum entropy sampling and optimal Bayesian experimental design},
  author={Sebastiani, Paola and Wynn, Henry P},
  journal={Journal of the Royal Statistical Society: Series B (Statistical Methodology)},
  volume={62},
  number={1},
  pages={145--157},
  year={2000},
  publisher={Wiley Online Library}
}

@article{jiang2016quantification,
  title={Quantification of cell size using temporal diffusion spectroscopy},
  author={Jiang, Xiaoyu and Li, Hua and Xie, Jingping and Zhao, Ping and Gore, John C and Xu, Junzhong},
  journal={Magnetic resonance in medicine},
  volume={75},
  number={3},
  pages={1076--1085},
  year={2016},
  publisher={Wiley Online Library}
}

@article{dai2026optimizing,
  title={Optimizing IMPULSED Acquisition Protocols for Clinical 3T Scanners Through Bayesian Experimental Design},
  author={Dai, Yan and Jia, Xun and Aguilera, Todd and Jiang, Kai and Rodriguez, Arely Perez and Vanhaezebrouck, Isabelle and Deng, Jie},
  journal={arXiv preprint arXiv:2604.06020},
  year={2026}
}

@article{zhao2018optimal,
  title={Optimal experiment design for magnetic resonance fingerprinting: Cram{\'e}r-Rao bound meets spin dynamics},
  author={Zhao, Bo and Haldar, Justin P and Liao, Congyu and Ma, Dan and Jiang, Yun and Griswold, Mark A and Setsompop, Kawin and Wald, Lawrence L},
  journal={IEEE transactions on medical imaging},
  volume={38},
  number={3},
  pages={844--861},
  year={2018},
  publisher={IEEE}
}

@article{scope2022efficient,
  title={An efficient approach to optimal experimental design for magnetic resonance fingerprinting with B-splines},
  author={Scope Crafts, Evan and Lu, Hengfa and Ye, Huihui and Wald, Lawrence L and Zhao, Bo},
  journal={Magnetic resonance in medicine},
  volume={88},
  number={1},
  pages={239--253},
  year={2022},
  publisher={Wiley Online Library}
}

@article{uhl2025explainable,
    author = {Uhl, Quentin and Pavan, Tommaso and Gerold, Julianna and Chan, Kwok-Shing and Jun, Yohan and Fujita, Shohei and Bhatt, Aneri and Ma, Yixin and Wang, Qiaochu and Lee, Hong-Hsi and Huang, Susie Y. and Bilgic, Berkin and Jelescu, Ileana},
    title = {Reduced NEXI protocol for the quantification of human gray matter microstructure on the Connectome 2.0 scanner},
    journal = {Imaging Neuroscience},
    volume = {4},
    pages = {IMAG.a.1380},
    year = {2026},
    month = {10},
    issn = {2837-6056},
    doi = {10.1162/IMAG.a.1380},
}

@article{karger1985nmr,
  title={NMR self-diffusion studies in heterogeneous systems},
  author={K{\"a}rger, J{\"o}rg},
  journal={Advances in Colloid and Interface Science},
  volume={23},
  pages={129--148},
  year={1985},
  publisher={Elsevier}
}

@article{huang2021connectome,
  title={Connectome 2.0: Developing the next-generation ultra-high gradient strength human MRI scanner for bridging studies of the micro-, meso-and macro-connectome},
  author={Huang, Susie Y and Witzel, Thomas and Keil, Boris and Scholz, Alina and Davids, Mathias and Dietz, Peter and Rummert, Elmar and Ramb, Rebecca and Kirsch, John E and Yendiki, Anastasia and others},
  journal={Neuroimage},
  volume={243},
  pages={118530},
  year={2021},
  publisher={Elsevier}
}

@article{ramos2026ultra,
  title={Ultra-high gradient connectomics and microstructure MRI scanner for imaging of human brain circuits across scales},
  author={Ramos-Llord{\'e}n, Gabriel and Lee, Hong-Hsi and Davids, Mathias and Dietz, Peter and Krug, Andreas and Kirsch, John E and Mahmutovic, Mirsad and M{\"u}ller, Alina and Ma, Yixin and Lee, Hansol and others},
  journal={Nature biomedical engineering},
  volume={10},
  number={2},
  pages={309--324},
  year={2026},
  publisher={Nature Publishing Group UK London}
}

@article{rafael2020robust,
  title={Robust Monte-Carlo simulations in diffusion-MRI: Effect of the substrate complexity and parameter choice on the reproducibility of results},
  author={Rafael-Patino, Jonathan and Romascano, David and Ramirez-Manzanares, Alonso and Canales-Rodr{\'\i}guez, Erick Jorge and Girard, Gabriel and Thiran, Jean-Philippe},
  journal={Frontiers in neuroinformatics},
  volume={14},
  pages={8},
  year={2020},
  publisher={Frontiers Media SA}
}

@article{canales2025diffusion,
  title={A diffusion MRI model for random walks confined on cylindrical surfaces: towards non-invasive quantification of myelin sheath radius},
  author={Canales-Rodr{\'\i}guez, Erick J and Tax, Chantal MW and Fischi-Gomez, Elda and Jones, Derek K and Thiran, Jean-Philippe and Rafael-Pati{\~n}o, Jonathan},
  journal={Frontiers in Physics},
  volume={13},
  pages={1516630},
  year={2025},
  publisher={Frontiers Media SA}
}

@article{tournier2007robust,
  title={Robust determination of the fibre orientation distribution in diffusion MRI: non-negativity constrained super-resolved spherical deconvolution},
  author={Tournier, J-Donald and Calamante, Fernando and Connelly, Alan},
  journal={Neuroimage},
  volume={35},
  number={4},
  pages={1459--1472},
  year={2007},
  publisher={Elsevier}
}

@article{jeurissen2014multi,
  title={Multi-tissue constrained spherical deconvolution for improved analysis of multi-shell diffusion MRI data},
  author={Jeurissen, Ben and Tournier, Jacques-Donald and Dhollander, Thijs and Connelly, Alan and Sijbers, Jan},
  journal={NeuroImage},
  volume={103},
  pages={411--426},
  year={2014},
  publisher={Elsevier}
}

@article{taskin2025spatially,
  title={Spatially regularized super-resolved constrained spherical deconvolution (SR2-CSD) of diffusion MRI data},
  author={Taskin, Ekin and Girard, Gabriel and Haro, Juan Luis Villarreal and Rafael-Pati{\~n}o, Jonathan and Garyfallidis, Eleftherios and Thiran, Jean-Philippe and Canales-Rodr{\'\i}guez, Erick Jorge},
  journal={NeuroImage},
  pages={121656},
  year={2025},
  publisher={Elsevier}
}

@article{veraart2020noninvasive,
  title={Noninvasive quantification of axon radii using diffusion MRI},
  author={Veraart, Jelle and Nunes, Daniel and Rudrapatna, Umesh and Fieremans, Els and Jones, Derek K and Novikov, Dmitry S and Shemesh, Noam},
  journal={elife},
  volume={9},
  pages={e49855},
  year={2020},
  publisher={eLife Sciences Publications, Ltd}
}

@article{assaf2008axcaliber,
  title={AxCaliber: a method for measuring axon diameter distribution from diffusion MRI},
  author={Assaf, Yaniv and Blumenfeld-Katzir, Tamar and Yovel, Yossi and Basser, Peter J},
  journal={Magnetic Resonance in Medicine: An Official Journal of the International Society for Magnetic Resonance in Medicine},
  volume={59},
  number={6},
  pages={1347--1354},
  year={2008},
  publisher={Wiley Online Library}
}

@article{barakovic2023estimating,
  title={Estimating axon radius using diffusion-relaxation MRI: calibrating a surface-based relaxation model with histology},
  author={Barakovic, Muhamed and Pizzolato, Marco and Tax, Chantal MW and Rudrapatna, Umesh and Magon, Stefano and Dyrby, Tim B and Granziera, Cristina and Thiran, Jean-Philippe and Jones, Derek K and Canales-Rodr{\'\i}guez, Erick J},
  journal={Frontiers in neuroscience},
  volume={17},
  pages={1209521},
  year={2023},
  publisher={Frontiers Media SA}
}

\end{document}